\documentclass[12pt]{iopart}
\usepackage{graphicx,amssymb,color}

\begin{document}

\title[Magnetization and magnetocaloric data of spin-1/2 Heisenberg regular polyhedra]{Enhanced magnetocaloric effect 
in a proximity of magnetization steps and jumps of spin-1/2 XXZ Heisenberg regular polyhedra}

\author{Katar\'ina Kar\v{l}ov\'a$^1$, Jozef Stre\v{c}ka$^1$, Johannes Richter$^2$}

\address{$^1$ Institute of Physics, Faculty of Science, P. J. \v{S}af\'arik University, Park Angelinum 9, 040 01 Ko\v{s}ice, Slovakia}
\address{$^2$ Institut f\"ur Theoretische Physik, Otto-von-Guericke Universit\"at in Magdeburg, P.O. Box 4120, 39016 Magdeburg, Germany}
\ead{katarina.karlova@student.upjs.sk}

\begin{abstract}
The magnetization process and adiabatic demagnetization of the antiferromagnetic spin-1/2 XXZ Heisenberg clusters with the shape of regular polyhedra (tetrahedron, octahedron, cube, icosahedron and dodecahedron) are examined using the exact diagonalization method. It is demonstrated that a quantum ($xy$) part of the XXZ exchange interaction is a primary cause for presence of additional intermediate magnetization plateaux and steps, which are totally absent in the limiting Ising case. The only exception to this rule is the spin-1/2 XXZ Heisenberg tetrahedron, which shows 
just a quantitative shift of the level-crossing fields related to two magnetization steps. It is evidenced that the spin-1/2 XXZ Heisenberg regular polyhedra exhibit an enhanced magnetocaloric effect in a proximity of 
the magnetization steps and jumps, which are accompanied with a rapid drop (rise) of temperature just above (below) of the level-crossing field when the magnetic field is removed adiabatically.
\end{abstract}
\vspace{2pc}
\vspace{2pc}
\submitto{\JPCM}
%
%
%
%
%

\section{Introduction}
The magnetization curves of the quantum Heisenberg spin systems can exhibit a lot of intriguing features such as intermediate plateaux 
\cite{hon2004,lacr11}, quasi-plateaux \cite{bell14,ohan15}, the magnetization steps and 
jumps \cite{prl02,shap02}, the magnetization cusp \cite{hone03,kawa01,saka04}, ramp \cite{saka11,sakai11} or kink singularities \cite{meis07}. A great deal of attention has been recently paid to the magnetization process of the quantum Heisenberg spin 
clusters, which afford paradigmatic theoretical models for molecular-based magnetic materials with strong intra-molecular and weak inter-molecular interactions between the spin centers. The molecular magnets, which involve as a magnetic core relatively small spin 
clusters, are fascinating topic of current research interest due to a wide spectrum of possible applications ranging from condensed matter physics, magneto-chemistry, 
biology, biomedicine, material science up to quantum computing
\cite{bart14,schnack2010,scho04}. 

In the past, the particular attention has been devoted to the simplest geometrically frustrated Heisenberg spin clusters, which are composed of a few interacting magnetic centers as for instance the 
spin triangle \cite{hara05}, tetrahedron \cite{bose05}, 
cuboctahedron \cite{schmidt2005,schnack09,karl17,hone09,schnalle09}, truncated tetrahedron \cite{schnalle09,coff92} or even more complex clusters involving greater 
number of the interacting spin centers \cite{schnack2001,schn10,ioannis2008,kons09,axen01}. The effect of inter-molecular interactions 
between geometrically identical Heisenberg spin clusters has been also 
investigated in a variety of geometric spin arrangements including the coupled spin triangles \cite{schm10}, 
squares and cubes  \cite{schnack2016}, 
rings, icosahedrons \cite{schm01} or dodecahedrons \cite{kons15}. 

The Heisenberg spin clusters with a geometric shape of regular convex polyhedra (tetrahedron, octahedron, cube, icosahedron and dodecahedron) have the most beautiful magnetic architecture from the symmetry point of view. Namely, the regular polyhedra are the only (Platonic) solids constituted by the one and same regular polygon, which have all vertices, all edges and all faces completely 
equivalent (see Fig.~\ref{PlatSol}). Moreover, one may expect extraordinary rich magnetic behaviour of the regular Heisenberg polyhedra due to a mutual interplay of the geometric spin frustration and quantum effects, because only the spin cube is non-frustrated on assumption that the antiferromagnetic nearest-neighbour interaction is considered. 
Owing to this fact, the magnetization curves and magnetocaloric  properties of the 
regular Heisenberg polyhedra have been comprehensively studied \cite{kons05,huch11,kons14,schn06,kons07,park04}. Recently, it has been demonstrated 
that the regular Ising polyhedra also exhibit due to the geometric spin frustration several 
intriguing magnetic features such as presence of multiple intermediate magnetization plateaux 
and jumps in spite of the complete lack of quantum fluctuations \cite{voge93,stre15,karl16}. 
Thus, by variation of the corresponding anisotropy parameter
$\Delta$, see below, the strength of quantum fluctuations can be tuned
between the extreme quantum case (isotropic Heisenberg limit) and the
semi-classical case (Ising limit). Moreover, one may expect that the exchange 
anisotropy can be relevant for many magnetic molecules, but to the best of our knowledge, 
the effect of the exchange anisotropy has not been dealt with yet. 
In the present work, we will therefore fill in this gap by exploring the magnetization process and magnetocaloric properties of the spin-1/2 XXZ Heisenberg clusters with the shape of regular polyhedra. This study will clarify how the magnetic properties evolve with the exchange anisotropy from the isotropic Heisenberg limit up to the highly anisotropic Ising limit. 
A particular advantage of the investigation of the Platonic solids
presented in Fig.~\ref{PlatSol} consists in the fact that exact numerical data 
can be obtained for all physical quantities by using full numerical diagonalization.
Let us also mention that an enhanced magnetocaloric effect has been
recently reported for several molecular magnets \cite{schnack2007,zheng2011,sharples2013,sharples2014,fu2015}.
Hence, the finite-size magnetic spin clusters, especially if frustration is present,  
are promising candidates for future applications in magnetic refrigeration.       
  
The organization of this paper is as follows. The spin-1/2 XXZ Heisenberg regular polyhedra will be defined in Section \ref{model}, where a few details of the calculation procedure will be also presented. The most interesting results for the ground-state phase diagram, magnetization curves and magnetocaloric effect will be discussed in Section \ref{result}. Finally, our paper will end up in Section \ref{conclusion} with several concluding remarks. 

\section{Model and methods}
\label{model}

\begin{figure}[t]
\includegraphics[width=0.7\textwidth]{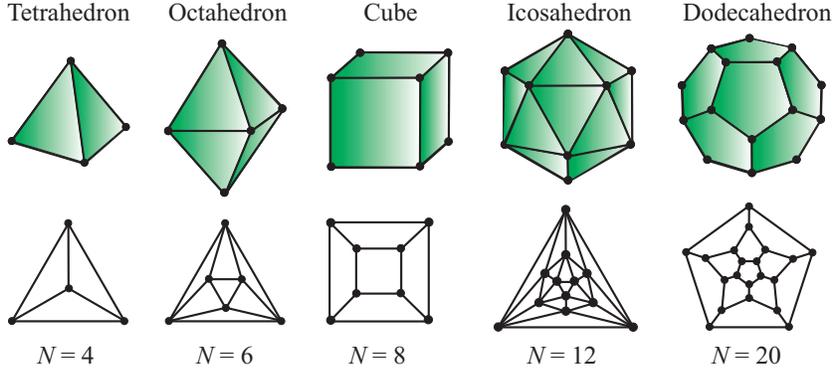}
\caption{(Color online) The five regular polyhedra (Platonic solids), which are composed from the unique regular polygon. The upper panel displays a steric arrangement of the regular polyhedra, while the lower panel shows the equivalent planar projections used for an enumeration of vertices. The number $N$ determines the total number of vertices of a given regular polyhedron.}
\label{PlatSol}
\end{figure}

Let us consider the spin-1/2 XXZ Heisenberg clusters with the shape of regular polyhedra (see Fig. \ref{PlatSol}) given by the Hamiltonian
\begin{eqnarray}
\hat{\cal H} = J \sum_{\langle i,j \rangle}^{N_b} \left[\Delta \left(\hat{S}_{i}^x\hat{S}_{j}^x + \hat{S}_{i}^y\hat{S}_{j}^y \right) + \hat{S}_{i}^z\hat{S}_{j}^z\right] - h \sum_{i=1}^N \hat{S}_i^z,
\label{ham}
\end{eqnarray}
where $\hat{\bf{S}}_{i} \equiv (\hat{S}_i^x, \hat{S}_i^y, \hat{S}_i^z)$ denotes the spatial components of the spin-1/2 operator placed at $i$th vertex of a regular polyhedron, the first summation accounts for the antiferromagnetic XXZ Heisenberg interaction $J>0$ between the nearest-neighbour spins and $\Delta \in \langle 0; 1\rangle$ is the (inverse) anisotropy parameter. Two limiting cases $\Delta=0$ and $\Delta=1$ accordingly correspond to the Ising and isotropic Heisenberg models, respectively. The second summation accounts for the Zeeman's energy of magnetic moments in the external magnetic field $h > 0$ and finally, $N$ ($N_b$) denotes the total number of vertices (edges) of a given regular polyhedron. 
 
Exact results for the spin-1/2 XXZ Heisenberg tetrahedron can be easily derived by Kambe projection method \cite{kambe50}, which employs a validity of the commutation relation between a square of the total spin operator $\hat{\bf{S}}_{T} = \sum_{i=1}^4 \hat{\bf{S}}_{i}$ and its $z$-component $\hat{S}_{T}^z = \sum_{i=1}^4 \hat{S}_{i}^z$ with the Hamiltonian (\ref{ham}), i.e. $\left[\hat{\cal H}, \hat{\bf{S}}_{T}^2 \right] = \left[\hat{\cal H}, \hat{S}_{T}^z \right]=0$. This means that the total spin and its $z$-component are conserved quantities with well defined quantum spin numbers $S_{T} = 0,1,2$ and $S_{T}^z = -S_{T}, -S_{T}+1, \ldots, S_{T}$. The energy eigenvalues can be then expressed in terms of the quantum spin numbers $S_{T}$ and $S_{T}^z$
\begin{eqnarray}
\fl E_{S_{T},S_{T}^z}=-\frac{J(1-\Delta)}{2}\left[S_{T}(S_{T}+1)-\left(S_{T}^z\right)^2-2\right]  + \frac{J}{2}\left[S_{T}(S_{T}+1)-3\right] -hS_{T}^z.
\label{energy}
\end{eqnarray}

The canonical partition function of the spin-1/2 XXZ Heisenberg tetrahedron is defined through the relation
\begin{equation}
{\cal Z} = \sum_{S_{T}} \sum_{S_{T}^z} g_{S_{T}} \exp (-\beta E_{S_{T},S_{T}^z}),
\label{zt}
\end{equation}
where $\beta = 1/(k_{\rm B} T)$, $k_{\rm B}$ is Boltzmann's constant, $T$ is the absolute temperature and $g_{S_{T}}$ denotes degeneracy of the energy levels $E_{S_{T},S_{T}^z}$. To obtain the degeneracy of the individual energy levels one may simply use a tensor product of the spin operators $\left( \frac{1}{2} \otimes \frac{1}{2} \right) \otimes \left( \frac{1}{2} \otimes \frac{1}{2} \right) = \left( 0 \oplus 1 \right) \otimes \left( 0 \oplus 1 \right) = 0 \oplus 1 \oplus 1 \oplus 0 \oplus 1 \oplus 2$, which implies the double degeneracy of the singlet state $S_{T}=0$, the triple degeneracy of the triplet state $S_{T}=1$ and the uniqueness of the quintuplet state $S_{T}=2$. Hence, one may exactly calculate the partition function (\ref{zt}) and the Gibbs free energy of the spin-1/2 XXZ Heisenberg tetrahedron
\begin{eqnarray}
G=\!&&\!-k_{\rm{B}}T\ln{\cal Z} = -k_{\rm{B}}T\ln \left\{2\exp\left[\beta J\left(\Delta + \frac{1}{2}\right)\right] \right.\nonumber \\ \!&&\! + 6\exp\left(\frac{1}{2}\beta J\Delta\right)\cosh\left(\beta h\right)+3\exp\left(\frac{1}{2}\beta J\right) + 2\exp\left(-\frac{3}{2}\beta J \right)\cosh\left(2\beta h\right)\nonumber \\ \!&&\! \left. + 2\exp\left(-\frac{3}{2}\beta J \Delta\right)\cosh\left(\beta h\right)  + \exp\left[\beta J\left(-2\Delta + \frac{1}{2}\right)\right]\right\}.
\label{gfe} 
\end{eqnarray}
The closed-form expression (\ref{gfe}) for the Gibbs free energy allows a straightforward calculation of the magnetization and the entropy using the thermodynamic relations
\begin{equation}
m=-\left(\frac{\partial G}{\partial h}\right)_T, \qquad S=-\left(\frac{\partial G}{\partial T}\right)_h.
\label{m_a_S}
\end{equation}

\begin{figure}
\centering
\includegraphics[width=0.3\textwidth]{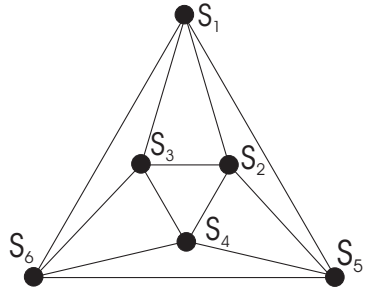}
\hspace{-0.5cm}
\caption{The Heisenberg octahedron with the notation for individual spins.}
\label{fig:Oktacik}
\end{figure}

Unfortunately, the aforedescribed procedure cannot be straightforwardly generalized in order to derive similar exact closed-form expressions for the other regular Heisenberg polyhedra. To illustrate the case, let us consider the spin-1/2 Heisenberg octahedron (Fig.~\ref{fig:Oktacik}) given by the Hamiltonian 
\begin{eqnarray}
\hat{\cal H} = J &(&\!\!\!\! \hat{\bf{S}}_{1}\cdot\hat{\bf{S}}_{2} + \hat{\bf{S}}_{1}\cdot\hat{\bf{S}}_{3} + \hat{\bf{S}}_{1}\cdot\hat{\bf{S}}_{5} + \hat{\bf{S}}_{1}\cdot\hat{\bf{S}}_{6} 
                 + \hat{\bf{S}}_{2}\cdot\hat{\bf{S}}_{3} + \hat{\bf{S}}_{2}\cdot\hat{\bf{S}}_{4} + \hat{\bf{S}}_{2}\cdot\hat{\bf{S}}_{5} \nonumber \\ 
								 &+&\hat{\bf{S}}_{3}\cdot\hat{\bf{S}}_{4} + \hat{\bf{S}}_{3}\cdot\hat{\bf{S}}_{6} + \hat{\bf{S}}_{4}\cdot\hat{\bf{S}}_{5} 
								 + \hat{\bf{S}}_{4}\cdot\hat{\bf{S}}_{6} + \hat{\bf{S}}_{5}\cdot\hat{\bf{S}}_{6})	- h \sum_{i=1}^6 \hat{S}_i^z,
\label{hamo}
\end{eqnarray}
which can be fully diagonalized by the Kambe projection method \cite{kambe50} only if the isotropic coupling is considered. At first, let us introduce the total spin operator $\hat{\bf{S}}_{T}$ and three composite spin operators $\hat{\bf{S}}_{14}$, $\hat{\bf{S}}_{26}$ and $\hat{\bf{S}}_{35}$  
\begin{eqnarray}
\hat{\bf{S}}_{T} = \sum_{i=1}^6 \hat{\bf{S}}_{i}, \quad 
\hat{\bf{S}}_{14} =  \hat {\bf{S}}_{1} + \hat{\bf{S}}_{4}, \quad 
\hat{\bf{S}}_{26} = \hat {\bf{S}}_{2} + \hat{\bf{S}}_{6}, \quad 
\hat{\bf{S}}_{35} =  \hat {\bf{S}}_{3} + \hat{\bf{S}}_{5},
\label{tridvojice}
\end{eqnarray}
which will serve for a diagonalization of the Hamiltonian (\ref{hamo}) of the spin-1/2 Heisenberg octahedron. The individual spins are enumerated according to the notation presented in Fig.~\ref{fig:Oktacik}, whereas the composite spin operators are always formed by two non-interacting spins located in opposite corners of the regular octahedron. It can be easily verified that the Hamiltonian (\ref{hamo}) 
commutes with a square of all spin operators (\ref{tridvojice}) as well as $z$-component 
of the total spin operator, i.e. $\left[\hat{\cal H}, \hat{\bf{S}}_{T}^2 \right] = \left[\hat{\cal H}, \hat{\bf{S}}_{14}^2 \right] =\left[\hat{\cal H}, \hat{\bf{S}}_{26}^2 \right] = \left[\hat{\cal H}, \hat{\bf{S}}_{35}^2 \right] = \left[\hat{\cal H}, \hat{S}_{T}^z \right]=0$. This fact allows us to express the energy eigenvalues 
in terms of the eigenvalues of the total and composite spin operators 
\begin{eqnarray}
\fl E_{S_{T}, S_{14}, S_{35}, S_{26}, S_{T}^z} = \frac{J}{2}  \left[S_{T}(S_{T} + 1) - S_{14}(S_{14} + 1) - S_{26}(S_{26} + 1) - S_{35}(S_{35} + 1) \right] - h S_{T}^z, \nonumber \\
\label{HEokta}
\end{eqnarray}
which can in principle acquire the following values $S_{T}=0,1,2,3$; $S_{14}=0,1$; $S_{26}=0,1$; $S_{35}=0,1$; $S_{T}^z=-S_{T},-S_{T}+1,\ldots,S_{T}$. To obtain the allowed combinations of the quantum spin numbers $S_{T}$, $S_{14}$, $S_{26}$ and $S_{35}$, which will also determine the respective degeneracy of individual energy levels, one may repeatedly use with a success the tensor product 
of the spin operators 
\begin{eqnarray}
\fl && \left( \frac{1}{2} \otimes \frac{1}{2} \right)  \otimes \left( \frac{1}{2} \otimes \frac{1}{2} \right) \otimes \left( \frac{1}{2} \otimes \frac{1}{2} \right) = \left( 0 \oplus 1 \right) \otimes \left( 0 \oplus 1 \right) \otimes \left( 0 \oplus 1 \right) \nonumber \\ 
\fl &=& 0 \oplus 1 \oplus 1 \oplus 0 \oplus 1 \oplus 2  \oplus 1 \oplus 0 \oplus 1 \oplus 2  \oplus 0 \oplus 1 \oplus 1 \oplus 0 \oplus 1 \oplus 2 \oplus 2 \oplus 1 \oplus 2 \oplus 3.
\label{oktaTenzor}
\end{eqnarray}
The exact result for the Gibbs free energy of the spin-1/2 Heisenberg octahedron obtained after taking into account all allowed combinations of the quantum spin numbers $S_{T}$, $S_{14}$, $S_{26}$, $S_{35}$ and $S_{T}^z$ reads as follows 
\begin{eqnarray}
G &=& -k_{\rm{B}}T\ln\{6 + 3\exp(\beta J) + 6\exp(2\beta J) + 4\exp(-3\beta J)  \nonumber \\ \!\!\!\!\!\!&&\!\!\!\!\!\! + \exp(3\beta J)  + 10\cosh(\beta h) + 4\cosh(2 \beta h) + 6\exp(\beta J)\cosh(\beta h) \nonumber \\ \!\!\!\!\!\!&&\!\!\!\!\!\! + 6\exp(-\beta J)\cosh(2\beta h)  + 2\exp(-3\beta J)\cosh(3\beta h) \nonumber \\ \!\!\!\!\!\!&&\!\!\!\!\!\! + 2\exp(-3\beta J)\cosh(2 \beta h)  + 2\exp(-3\beta J)\cosh(\beta h) \nonumber \\ \!\!\!\!\!\!&&\!\!\!\!\!\! + 6\exp(-\beta J)\cosh(\beta h) + 6\exp(2\beta J)\cosh(\beta h)\}.
\label{HZokta}
\end{eqnarray}
The magnetization and entropy of the spin-1/2 Heisenberg octahedron can be consequently obtained from Eq. (\ref{m_a_S}). It should be pointed out that this calculation procedure cannot be adapted neither to the spin-1/2 XXZ Heisenberg octahedron nor the spin-1/2 XXZ Heisenberg cube, icosahedron or dodecahedron.

To obtain the magnetization and entropy of the spin-1/2 XXZ Heisenberg octahedron, cube and icosahedron we have therefore performed the exact numerical diagonalization of the corresponding Hamiltonian (\ref{ham}) by adopting the subroutine edfulldiag from Algorithms and Libraries for Physics Simulations (ALPS) project \cite{bau11}. This approach allows a rigorous calculation of the magnetization and entropy data, which will be the central issue of our subsequent analysis focused on the magnetization process and magnetocaloric properties. The subroutine for exact diagonalization from ALPS project
provides equally-spaced numerical data that are suitable for a creation of the relevant density plots. On the other hand, the exact numerical diagonalization data for the spin-1/2 XXZ Heisenberg dodecahedron require application of a more sophisticated diagonalization code, since the size of the Hamiltonian matrix for $N=20$ is $1048576 \times 1048576$. Owing to this fact, the {\it Spinpack} package \cite{rich10} that exploits symmetries has been adapted in order to obtain the full spectrum of eigenvalues by a diagonalization of the Hamiltonian matrix of the Heisenberg dodecahedron in the set of Ising product states.

\section{Results and discussion}
\label{result}

In this section we will address the question how the magnetic behaviour of the spin-1/2 XXZ Heisenberg regular polyhedra with the antiferromagnetic nearest-neighbour interaction $J>0$ depends on the exchange anisotropy ranging from the extreme Ising limit $\Delta = 0$ up to the isotropic Heisenberg limit $\Delta = 1$. To clarify this issue, we will present exact analytical or numerical results for the ground-state phase diagrams, isothermal magnetization curves and adiabatic demagnetization of the spin-1/2 XXZ Heisenberg regular polyhedra. It is worthwhile to remark that all considered Heisenberg spin clusters are relatively small finite-size systems, which consequently have a discrete energy spectrum that is subject to Zeeman's splitting caused by the external magnetic field. It is therefore plausible to expect a crossing of 
the lowest-energy levels from different sectors with the $z$-component of the total spin $S_T^z$, which will have significant impact on a low-temperature 
magnetization process and thermodynamics. If the crossing of the energy levels from adjacent sectors with the 
total spin $S_T^z$ and $S_T^z+1$ takes place, then, the zero-temperature magnetization curve will exhibit at the respective level-crossing field the \textit{magnetization step} with the smallest possible change of the $z$-component of the total spin $\delta S_T^z = 1$. On the other hand, the zero-temperature magnetization curve will involve the appreciable \textit{magnetization jump} provided that the energy levels with a greater difference of the $z$-component of the total spin $\delta S_T^z > 1$ cross each other at a given level-crossing field.

\subsection{Spin-1/2 XXZ Heisenberg tetrahedron}
\begin{figure}[t]
\includegraphics[width=0.52\textwidth]{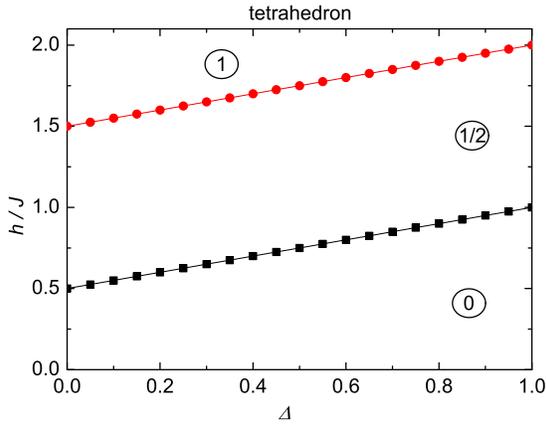}
\vspace*{-0.9cm}
\caption{The ground-state phase diagram of the spin-$1/2$ XXZ Heisenberg tetrahedron in the $\Delta-h/J$ plane.
 The acronyms
 determine the magnetization of a given lowest-energy 
eigenstate normalized with respect to its saturation value.}
\label{tetrF}
\end{figure}

Let us start our discussion with the ground-state phase diagram of the spin-1/2 XXZ Heisenberg tetrahedron, which is depicted in Fig.~\ref{tetrF} in the $\Delta - h/J$ plane. Three different areas delimited by the displayed level-crossing fields correspond to the 
lowest-energy eigenstates from the sectors with the total spin $S^z_T=0$, $1$ and $2$, whereas the corresponding value of the total magnetization normalized with respect to its saturation value is used for the notation purposes. It can be seen from Fig.~\ref{tetrF} that both level-crossing fields have an identical gradient in a respective linear dependence on the inverse anisotropy parameter $\Delta$. Owing to this fact, the width of zero magnetization plateau increases linearly with the parameter $\Delta$, while the width of intermediate one-half plateau remains constant.   
                                                                
\begin{figure}[t]
\includegraphics[width=0.52\textwidth]{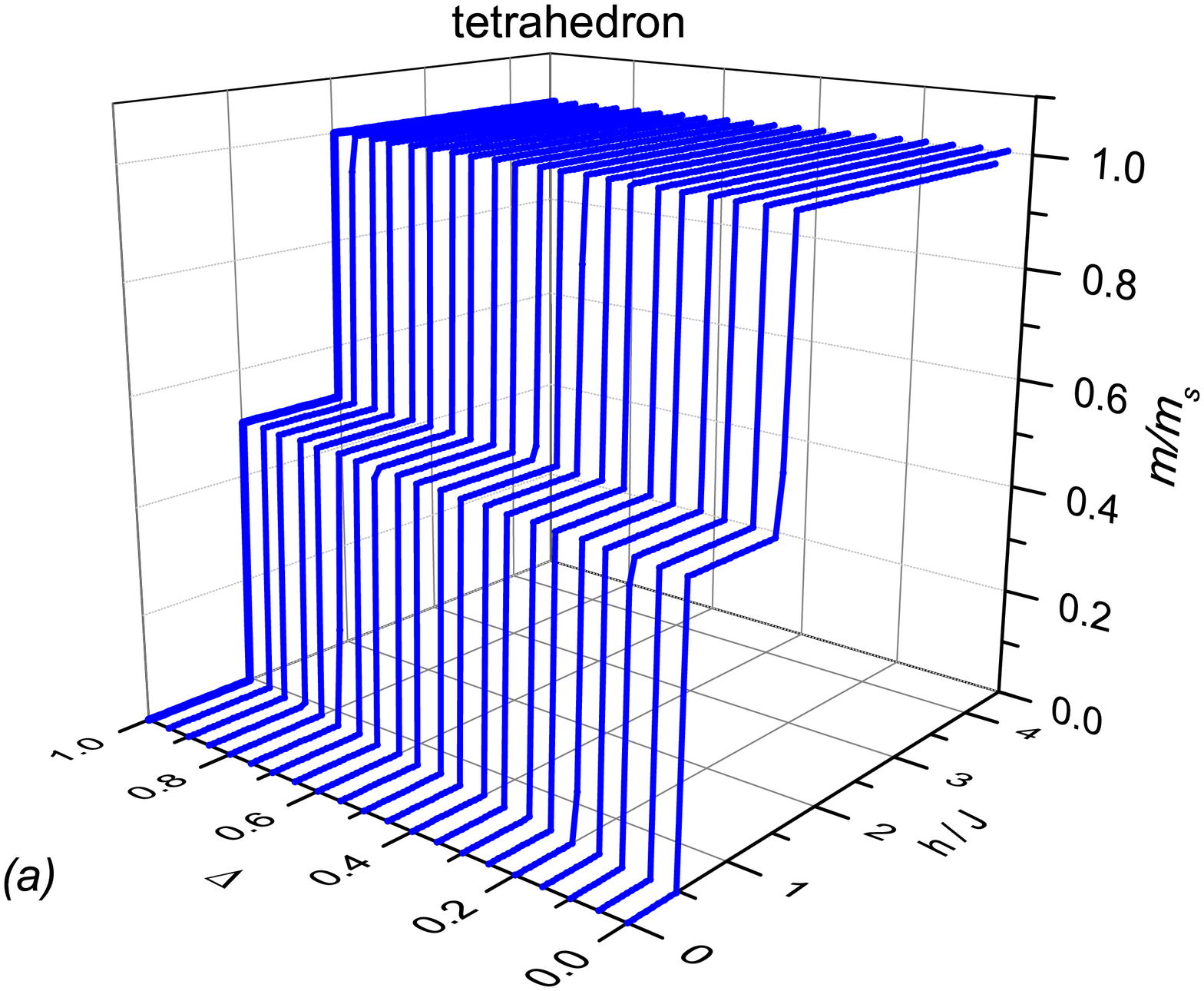}
\includegraphics[width=0.52\textwidth]{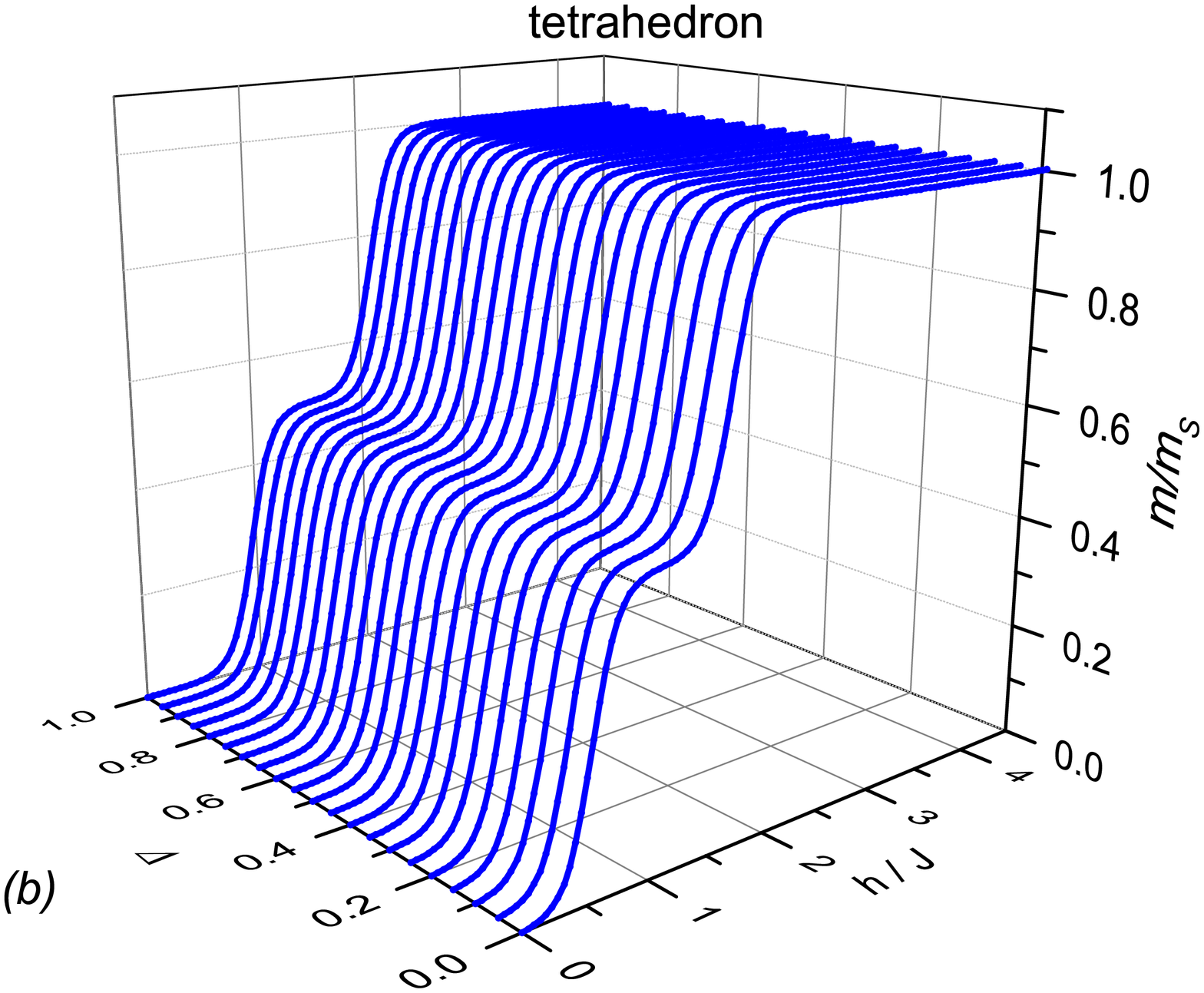}
\vspace*{-1.1cm}
\caption{The magnetization curves of the spin-$1/2$ XXZ Heisenberg tetrahedron for several values of the exchange anisotropy $\Delta$ 
and two different temperatures: (a) $k_{\rm{B}}T/J = 0.001$; (b) $k_{\rm{B}}T/J = 0.1$.}
\label{tetrM}
\end{figure}

To bring an insight into how the lowest-energy eigenstates are manifested in the isothermal magnetization curves we depict in Fig.~\ref{tetrM} the magnetization as a function of the magnetic field for several values of the anisotropy parameter $\Delta$ at the sufficiently low ($k_{\rm B} T/J = 0.001$) and moderate ($k_{\rm B} T/J = 0.1$) temperatures. Although there are no true magnetization plateaux and steps at any finite temperature, the low-temperature magnetization curve displayed in Fig.~\ref{tetrM}(a) is strongly reminiscent of the actual magnetization plateaux and steps 
observable strictly at zero temperature. Moreover, it can be seen from Fig.~\ref{tetrM}(a) that the magnetization plateaux at zero and one-half of the saturation magnetization are realized in the magnetization curve regardless of the anisotropy parameter $\Delta$. The one-half magnetization plateau is kept constant and the zero magnetization plateau becomes wider upon strengthening of $\Delta$ in agreement with the established ground-state phase diagram (Fig.~\ref{tetrF}). Besides, the magnetization data presented in Fig.~\ref{tetrM}(b) would suggest that the magnetization plateaux and steps are gradually smeared out at moderate temperatures. 

\begin{figure}[t]
\hspace{0.4cm}
\includegraphics[width=0.52\textwidth]{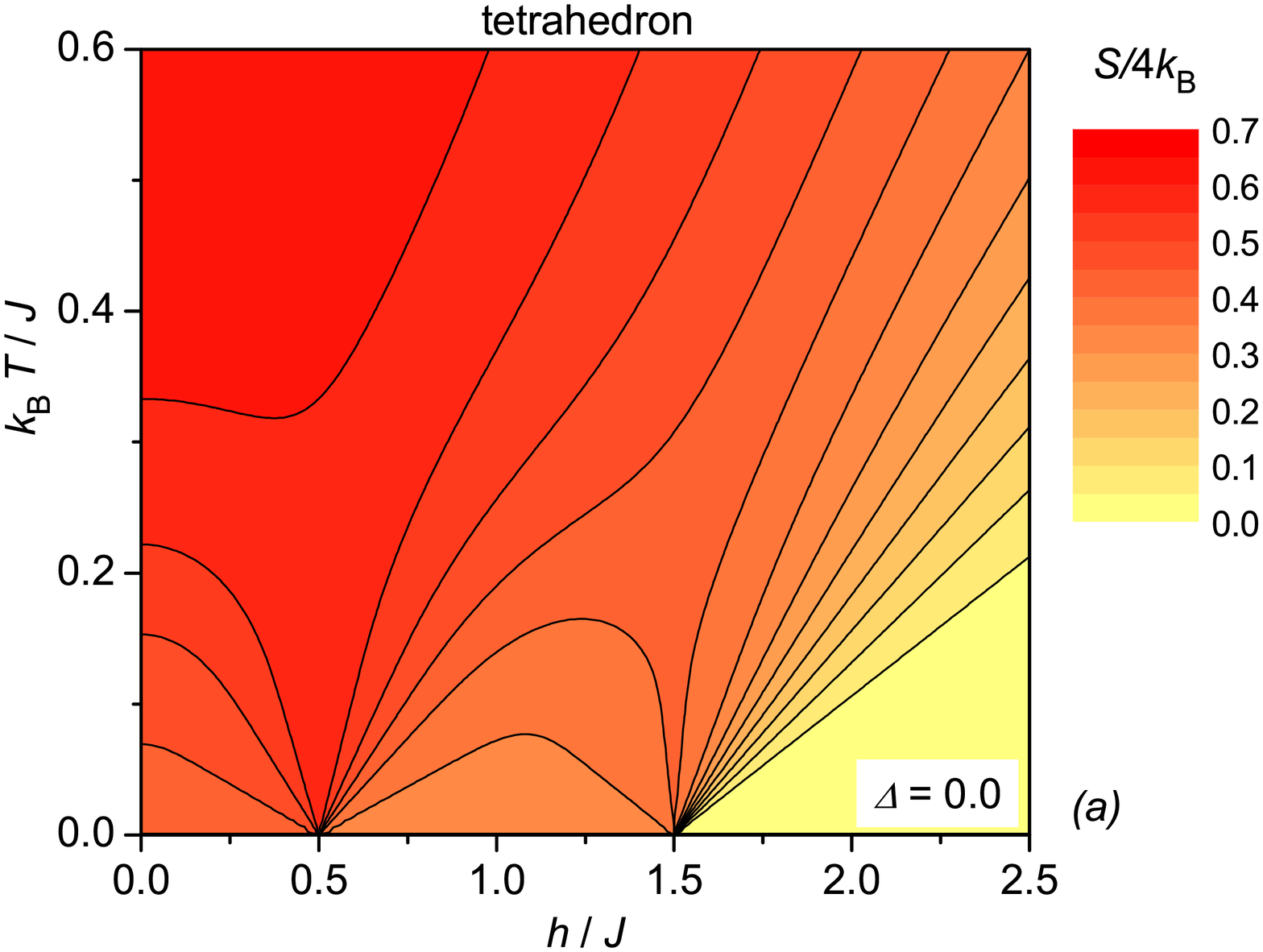}
\includegraphics[width=0.52\textwidth]{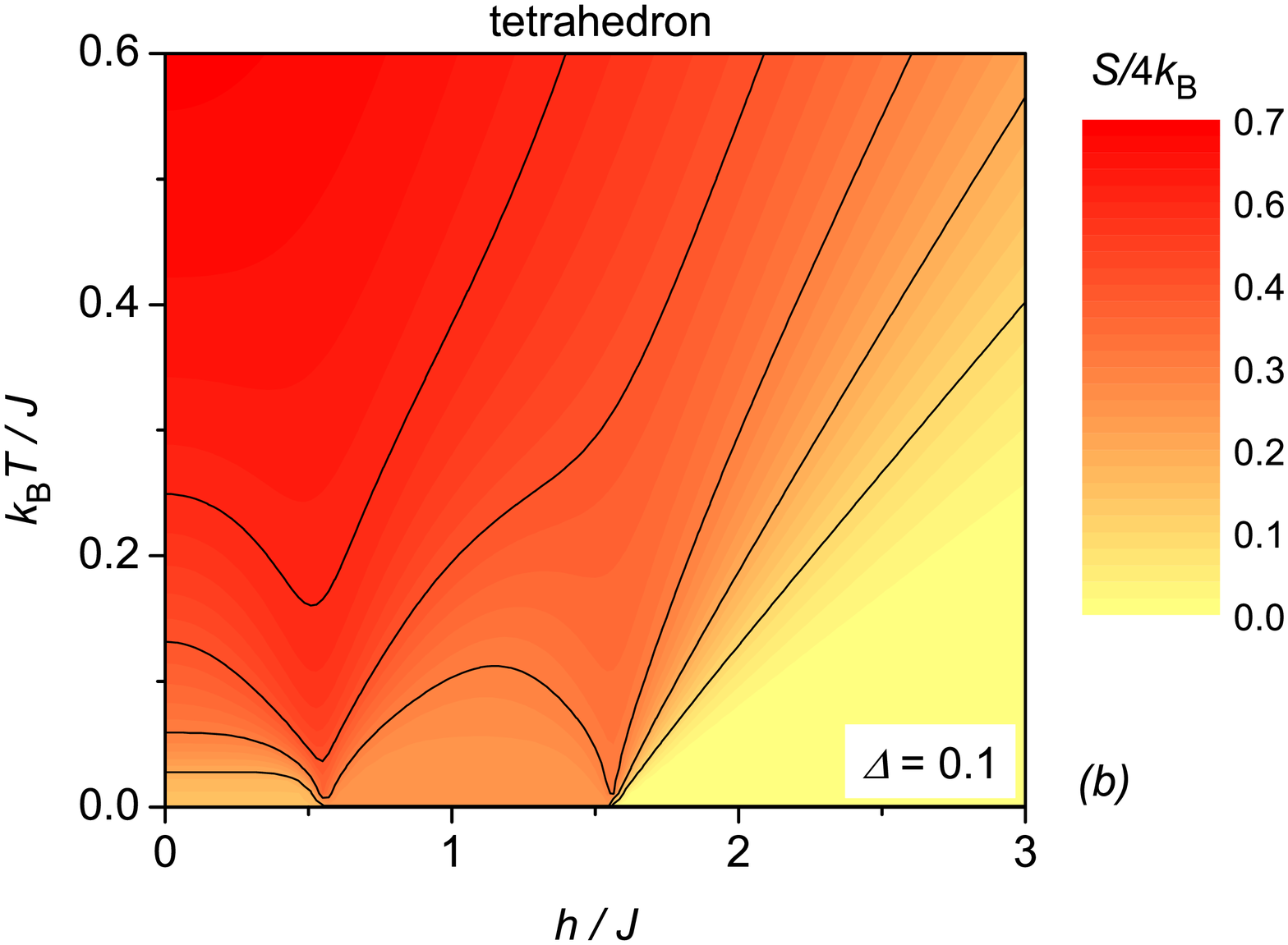}
\includegraphics[width=0.52\textwidth]{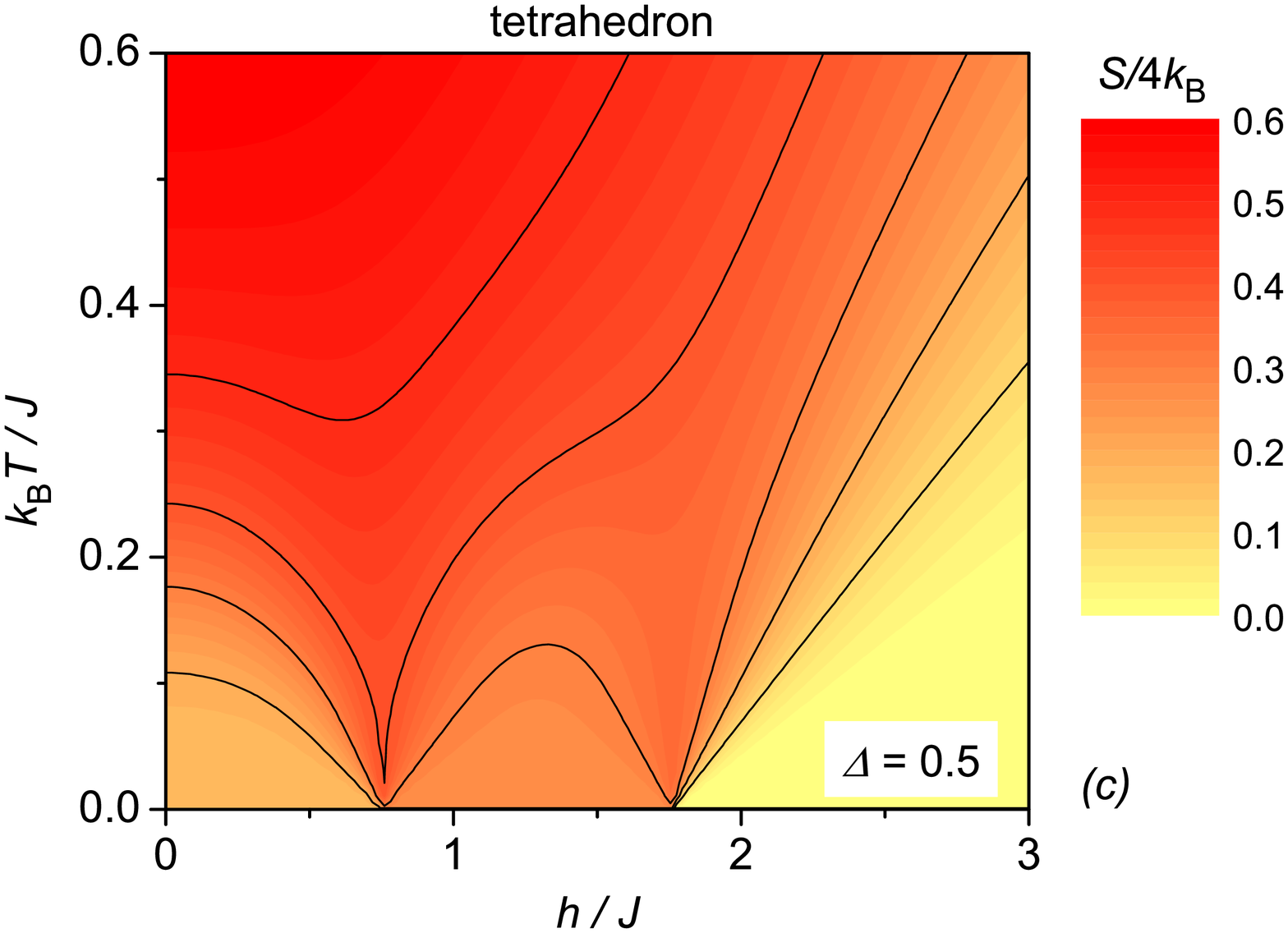}
\includegraphics[width=0.52\textwidth]{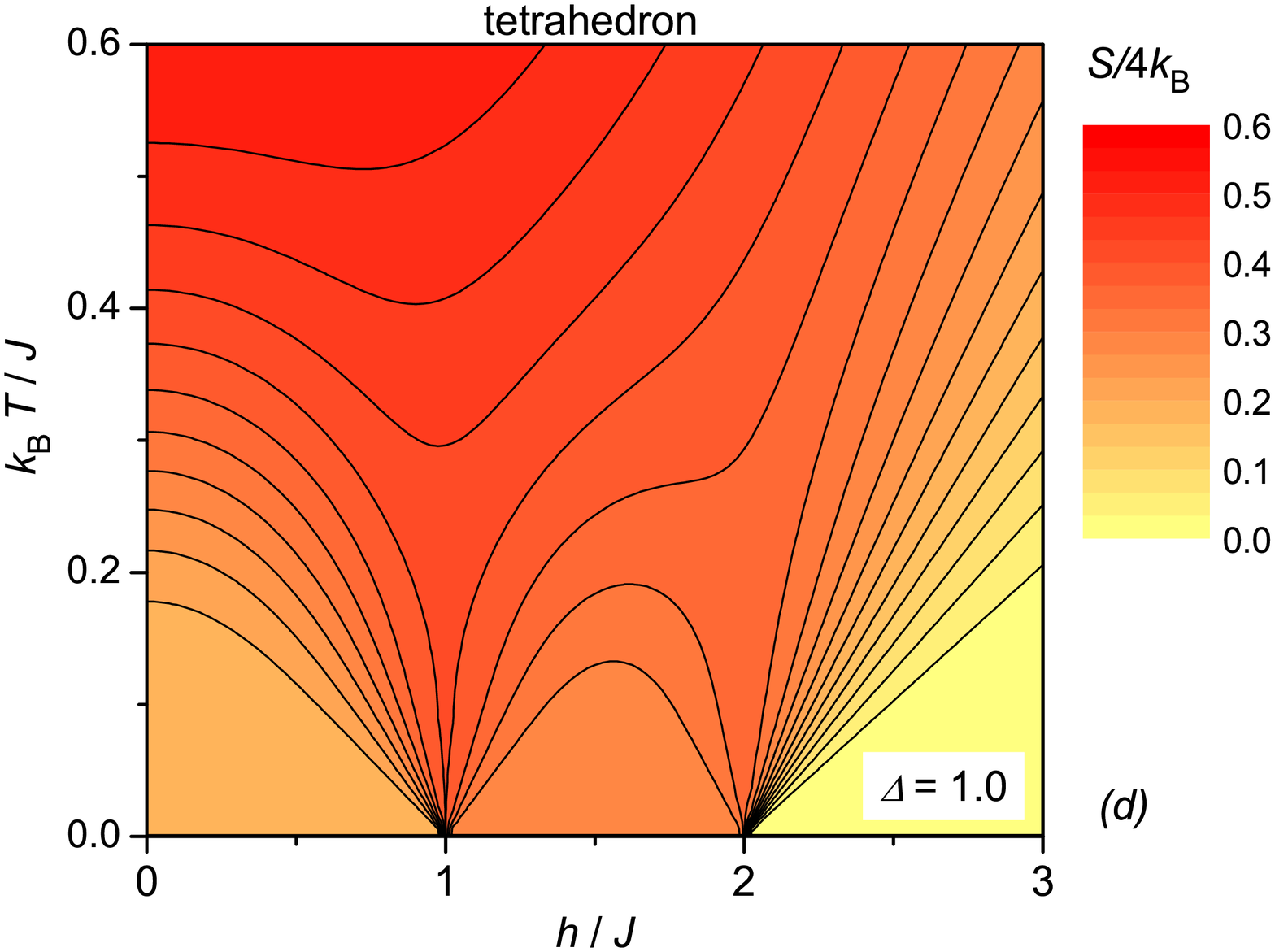}
\vspace*{-1.1cm}
\caption{A density plot of the entropy per spin of the spin-$1/2$ XXZ Heisenberg tetrahedron as a function of the magnetic field and temperature for four different values of the anisotropy parameter: (a) $\Delta=0.0$; (b) $\Delta=0.1$; (c) $\Delta=0.5$; (d) $\Delta=1.0$.}
\label{tetrE}
\end{figure}

Next, let us examine magnetocaloric properties of the spin-1/2 XXZ Heisenberg tetrahedron, which can be particularly interesting in a vicinity of the magnetization steps. A few isentropy lines are plotted in Fig.~\ref{tetrE}(a) in the field-temperature plane for the limiting Ising case with $\Delta=0$. The displayed isentropy lines can be alternatively viewed as the adiabatic temperature response achieved upon variation of the external magnetic field. The spin-1/2 Ising tetrahedron exhibits a giant magnetocaloric effect 
just above (below) of the level-crossing fields, where an abrupt drop (rise) of temperature is invoked upon lowering of the magnetic field. The qualitatively same dependences can be found also for the spin-1/2 XXZ Heisenberg tetrahedron regardless of whether one considers the strong easy-axis anisotropy $\Delta=0.1$ [Fig.~\ref{tetrE}(b)], the moderate easy-axis anisotropy $\Delta=0.5$ [Fig.~\ref{tetrE}(c)] or the completely isotropic coupling with $\Delta=1.0$ [Fig.~\ref{tetrE}(d)].

\subsection{Spin-1/2 XXZ Heisenberg octahedron}
\begin{figure}[t]
\includegraphics[width=0.52\textwidth]{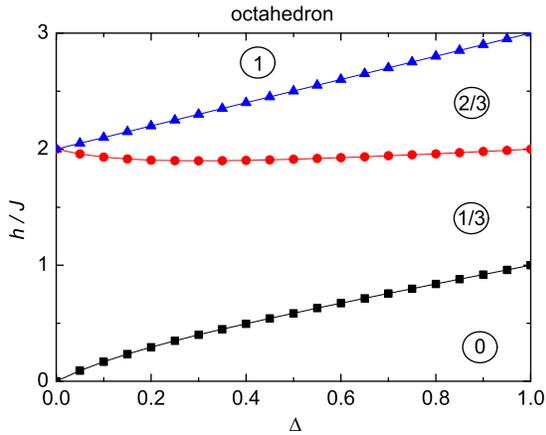}
\vspace*{-0.9cm}
\caption{The ground-state phase diagram of the spin-$1/2$ XXZ Heisenberg octahedron in the $\Delta-h/J$ plane. 
The acronyms 
 determine the magnetization of a given lowest-energy 
eigenstate normalized with respect to its saturation value.}
\label{octF}
\end{figure}

The ground-state phase diagram of the spin-1/2 XXZ Heisenberg octahedron is depicted in Fig.~\ref{octF} in the $\Delta - h/J$ plane. It is quite obvious from this figure that the zero-temperature magnetization curve of the spin-1/2 Ising octahedron exhibits an abrupt magnetization jump from the intermediate one-third plateau towards the full saturation. This is in sharp contrast to the zero-temperature magnetization curve of the spin-1/2 XXZ Heisenberg octahedron with arbitrary but non-zero inverse anisotropy $\Delta \neq 0$, which additionally displays two magnetization plateaux at zero and two-thirds of the saturation magnetization. Hence, it follows that the abrupt magnetization jump accompanied with the change of the total spin $\delta S_T^z = 2$ does merely occur in the Ising limit $\Delta = 0$, while the spin-1/2 XXZ Heisenberg octahedron with $\Delta > 0$ generally exhibits the smaller magnetization steps associated with the total spin change $\delta S_T^z = 1$. 

\begin{figure}[t]
\includegraphics[width=0.52\textwidth]{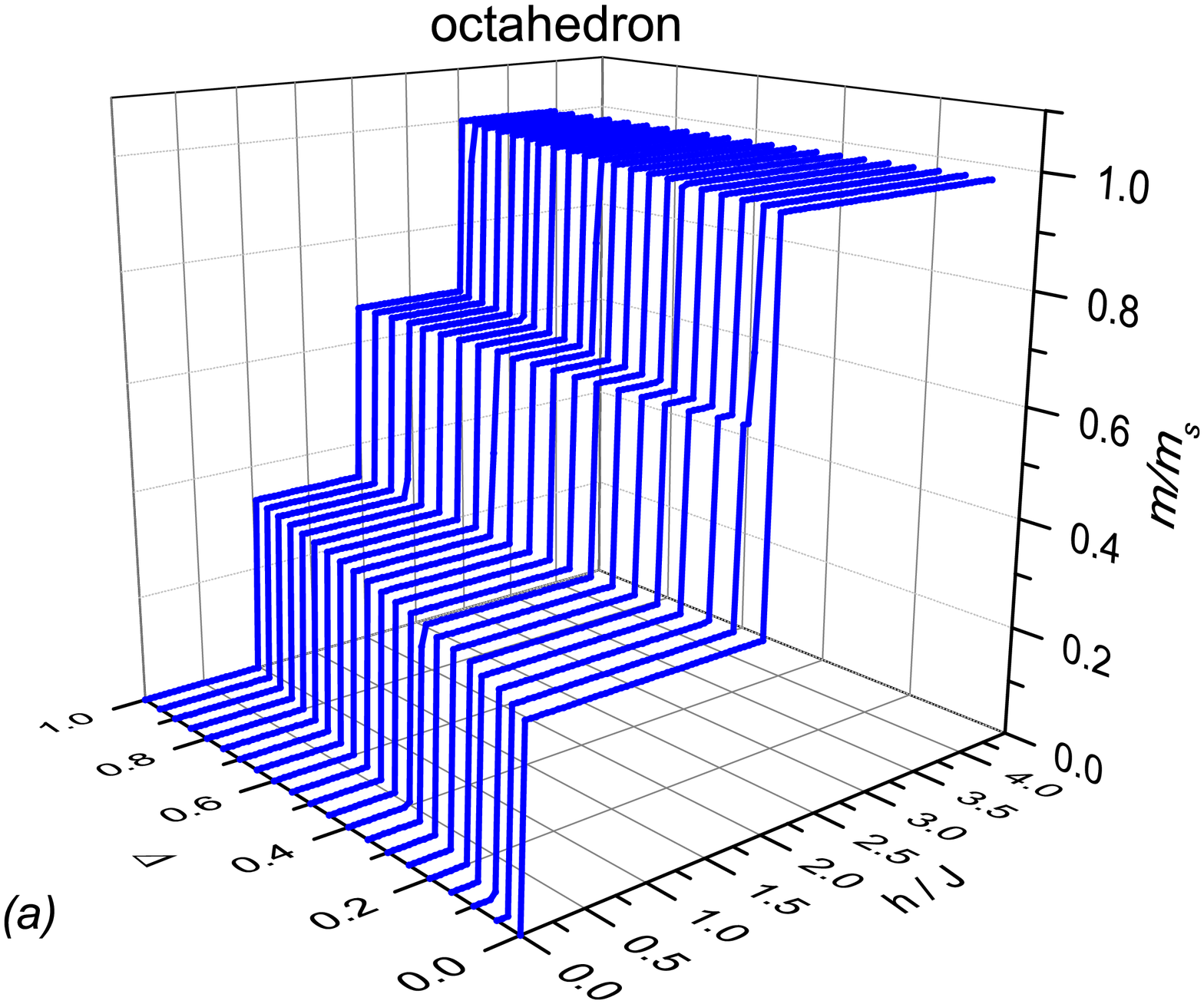}
\includegraphics[width=0.52\textwidth]{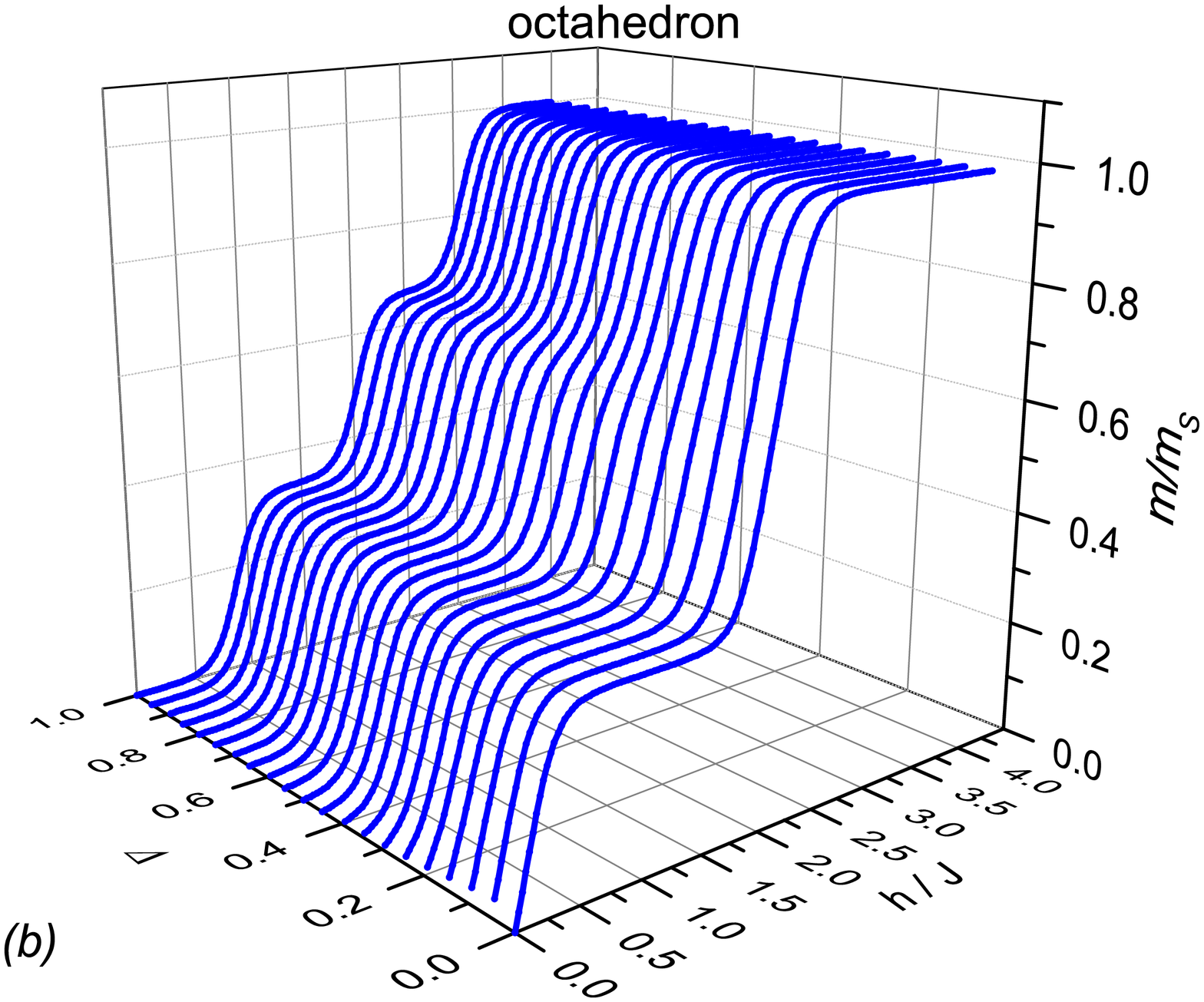}
\vspace*{-1.1cm}
\caption{The magnetization curves of the spin-$1/2$ XXZ Heisenberg octahedron for several values of the exchange anisotropy $\Delta$ 
and two different temperatures: (a) $k_{\rm{B}}T/J = 0.001$; (b) $k_{\rm{B}}T/J = 0.1$.}
\label{octM}
\end{figure}

To support this statement, we have depicted in Fig.~\ref{octM} the isothermal magnetization curves of the spin-1/2 XXZ Heisenberg octahedron for several values of the anisotropy parameter $\Delta$ and two different temperatures. As one can see, the spin-1/2 Ising octahedron indeed exhibits in a low-temperature magnetization curve [Fig.~\ref{octM}(a)] 
just the intermediate one-third plateau, which breaks down just at the saturation field where the magnetization jumps to its saturated value. It is evident from Fig.~\ref{octM}(a) that two novel magnetization plateaux at zero and two-thirds of the saturation magnetization of the spin-1/2 XXZ Heisenberg octahedron gradually extend over a wider range of the magnetic fields with increasing of a quantum ($xy$) part of the XXZ exchange interaction $\Delta$. As a matter of fact, the one-third plateau is roughly by $50\%$ narrower for the isotropic Heisenberg octahedron ($\Delta=1$) than for the Ising octahedron ($\Delta=0$). It is also noteworthy that subtle plateaux at zero and two-thirds of the saturation magnetization rapidly diminish due to a thermal activation of low-lying excited states if the strong easy-axis exchange anisotropy is considered. In fact, the  zero and two-thirds magnetization plateaux are almost indiscernible in the isothermal magnetization curves plotted in Fig.~\ref{octM}(b) at the moderate temperature $k_{\rm{B}}T/J = 0.1$ 
for strong enough easy-axis anisotropies $\Delta \lesssim 0.3$.
 
\begin{figure}[t]
\hspace{-0.5cm}
\includegraphics[width=0.52\textwidth]{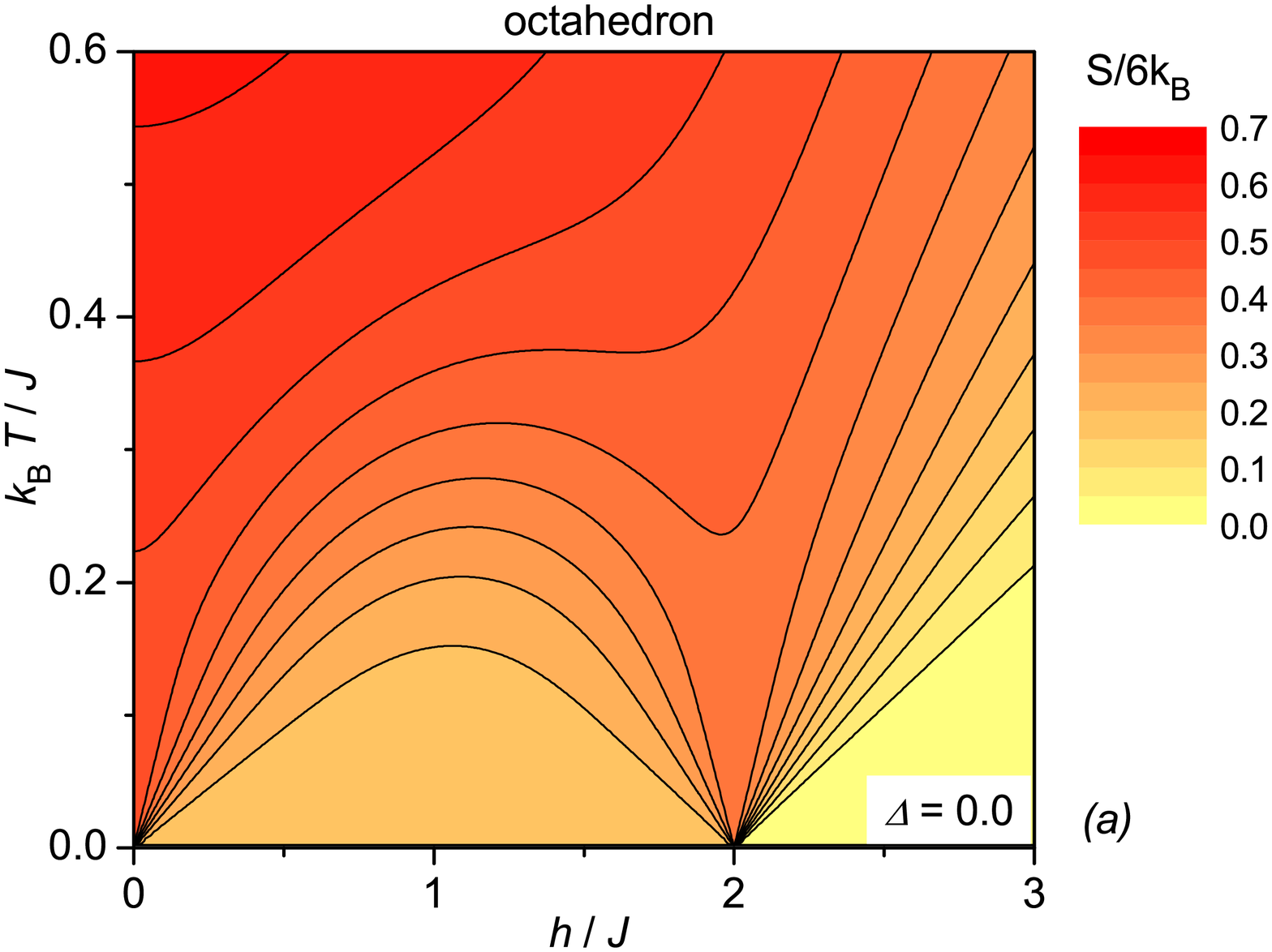}
\hspace{0.3cm}
\includegraphics[width=0.52\textwidth]{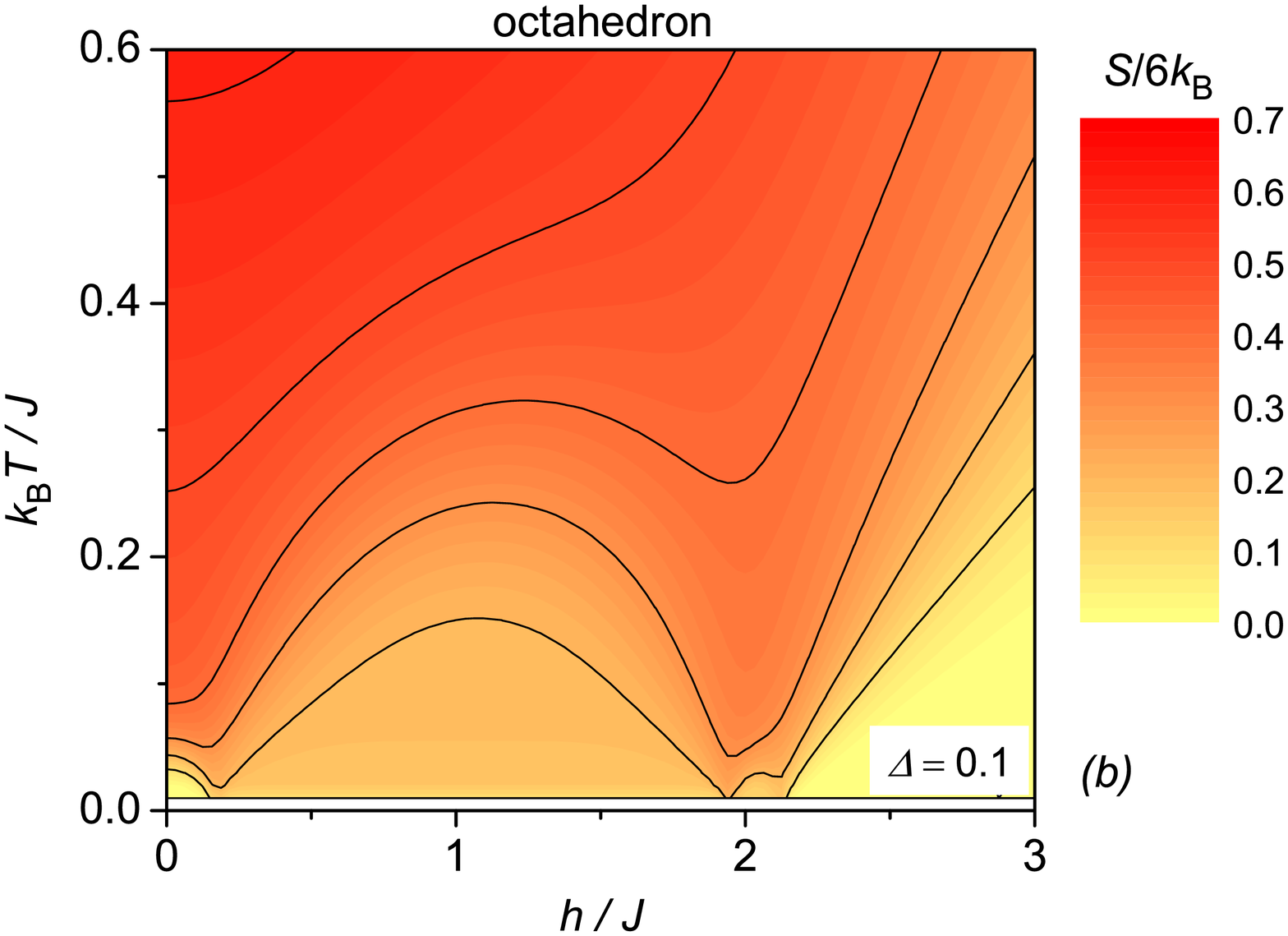}
\includegraphics[width=0.52\textwidth]{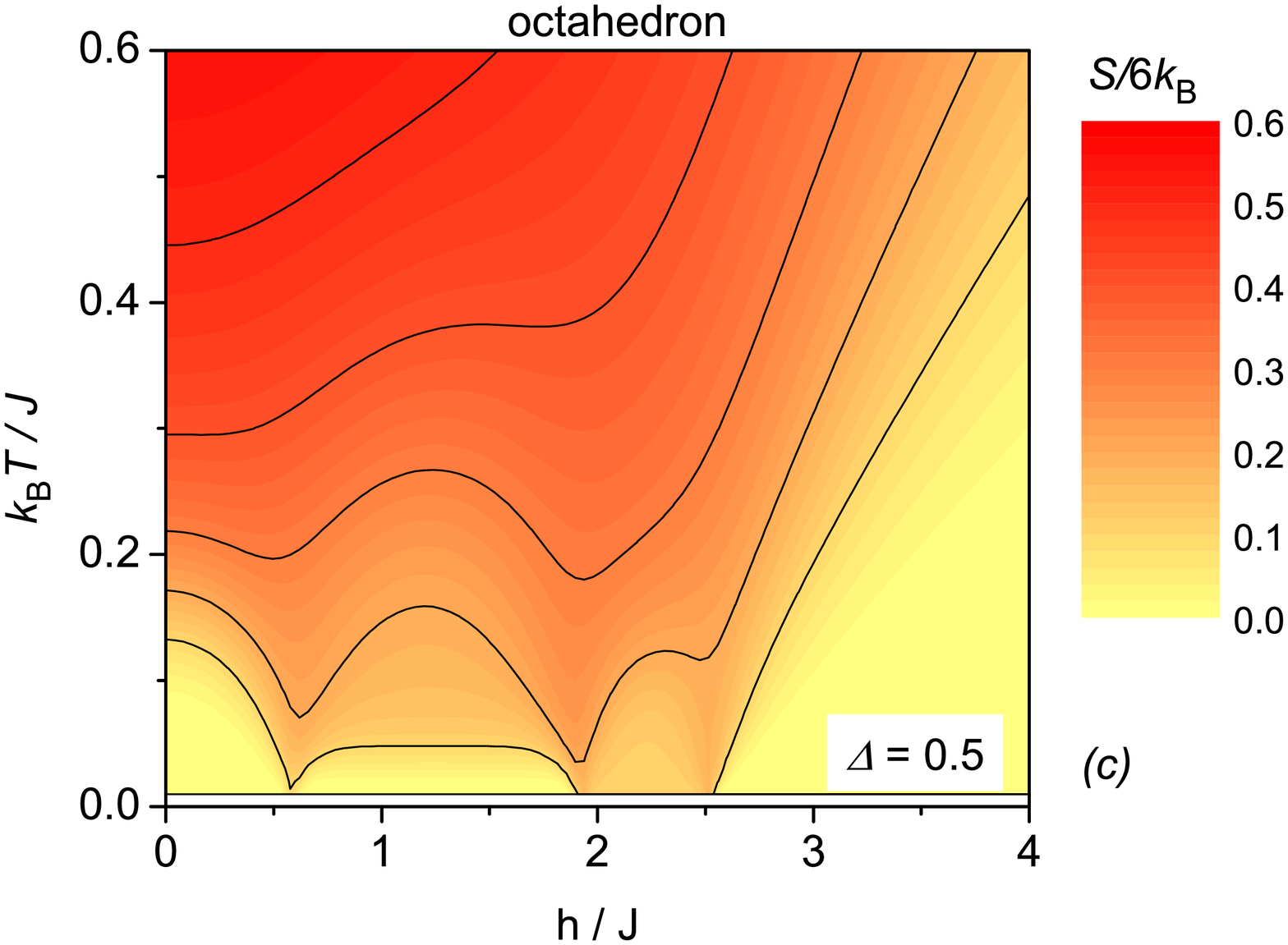}
\includegraphics[width=0.52\textwidth]{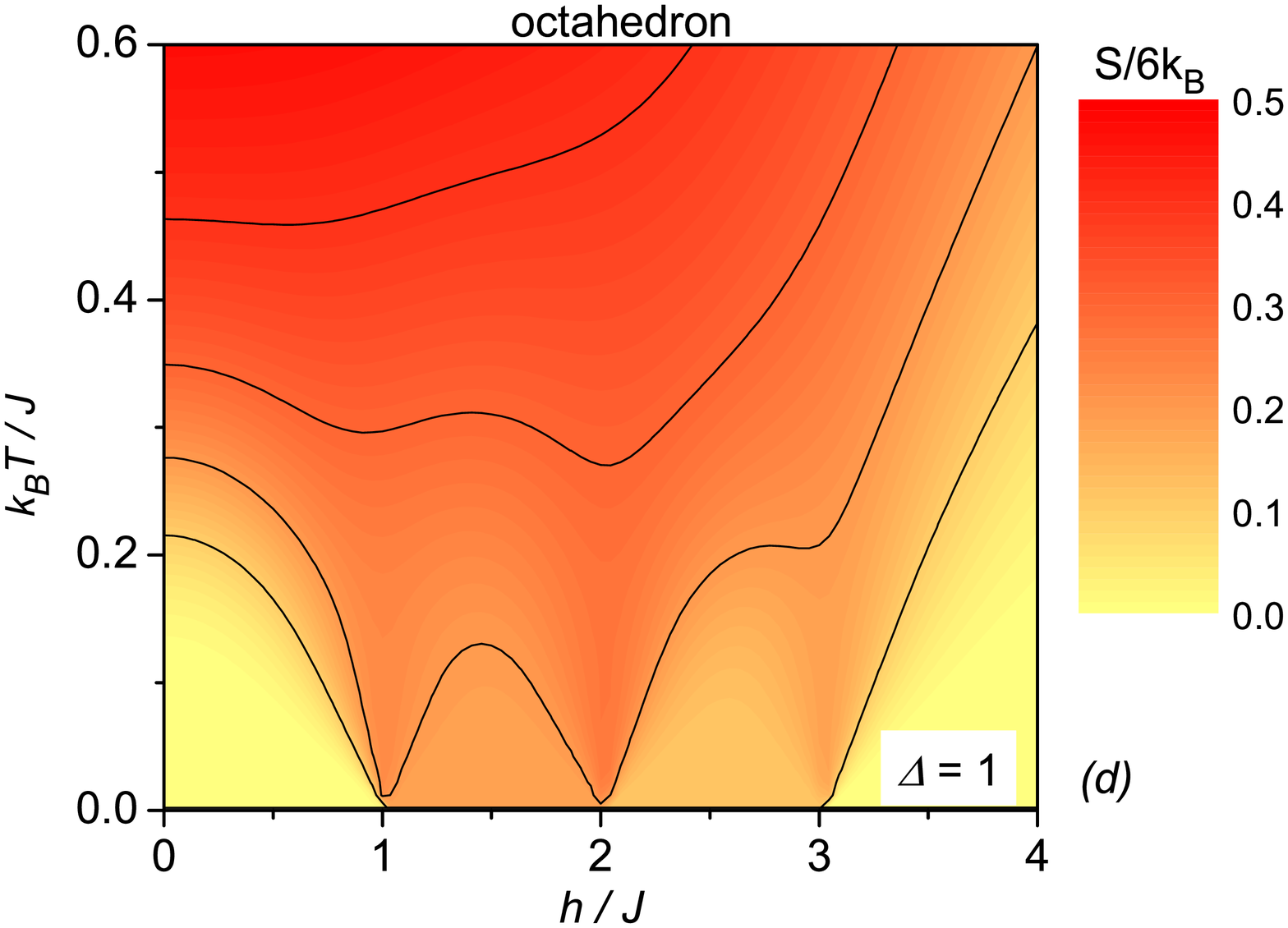}
\vspace*{-1.1cm}
\caption{A density plot of the entropy per spin of the spin-$1/2$ XXZ Heisenberg octahedron as a function of the magnetic field and temperature for four different values of the anisotropy parameter: (a) $\Delta=0.0$; (b) $\Delta=0.1$; (c) $\Delta=0.5$; (d) $\Delta=1.0$.}
\label{octE}
\end{figure}

It has been argued in our previous work \cite{stre15} that the absence of zero magnetization plateau in a low-temperature magnetization curve of the spin-1/2 Ising octahedron represents indispensable ground for observing a giant magnetocaloric effect in a proximity of zero magnetic field [see Fig.~\ref{octE}(a)], which makes this frustrated spin system quite promising refrigerant enabling cooling down to absolute zero temperature. Unfortunately, it will be shown hereafter that the quantum $xy$-part of the XXZ Heisenberg coupling prevents this intriguing magnetocaloric feature. Though it is still possible to observe under the adiabatic demagnetization of the spin-1/2 XXZ Heisenberg octahedron a relatively steep decrease (increase) of temperature just above (below) of each magnetization step [see Fig.~\ref{octE}(b)-(d)], however, temperature finally shows for any $\Delta \neq 0$ undesirable uprise to some finite value as the magnetic field approaches zero. The superior magnetocaloric effect can be thus detected 
just for the spin-1/2 XXZ Heisenberg octahedron with an extremely high easy-axis anisotropy $\Delta \lesssim 0.1$ [see e.g. Fig.~\ref{octE}(b)], where the undesirable deviation manifests itself only at very low temperatures due to a narrow field range corresponding to the zero magnetization plateau. 

\subsection{Spin-1/2 XXZ Heisenberg cube}
\begin{figure}[t]
\includegraphics[width=0.52\textwidth]{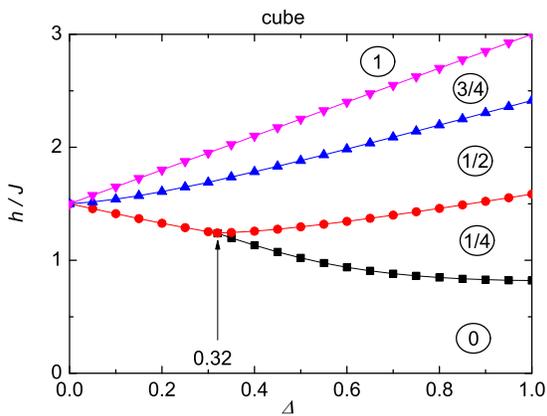}
\vspace*{-0.9cm}
\caption{The ground-state phase diagram of the spin-$1/2$ XXZ Heisenberg cube in the $\Delta-h/J$ plane. 
 The acronyms 
 determine the magnetization of a given lowest-energy 
eigenstate normalized with respect to its saturation value.}
\label{cubeF}
\end{figure}

The ground-state phase diagram of the antiferromagnetic spin-1/2 XXZ Heisenberg cube, which represents the only non-frustrated regular polyhedron, is displayed in Fig.~\ref{cubeF}. It is quite clear from this figure that the spin-1/2 Ising cube exhibits in the zero-temperature magnetization curve a direct magnetization jump from zero to the saturated value unlike the spin-1/2 XXZ Heisenberg cube, which may additionally display intermediate plateaux at one-quarter, one-half and three-quarters of the saturation magnetization. Interestingly, the one-quarter plateau is realized in the zero-temperature magnetization curve of the spin-1/2 XXZ Heisenberg cube 
just if $\Delta\gtrsim0.32$ in contrast to the one-half and three-quarters plateaux, which emerge 
for any $\Delta > 0$. This means that the magnetization curve of the spin-1/2 XXZ Heisenberg cube still involves for $\Delta\lesssim0.32$ one magnetization jump accompanied with the total spin change $\delta S_T^z = 2$ and other two magnetization steps due to a crossing of the energy levels with the smallest possible difference of the total spin $\delta S_T^z = 1$. On the other hand, the magnetization curve of the spin-1/2 XXZ Heisenberg cube includes for $\Delta\gtrsim0.32$ 
just four smaller magnetization steps associated with the total spin change $\delta S_T^z = 1$. 

\begin{figure}[t]
\includegraphics[width=0.52\textwidth]{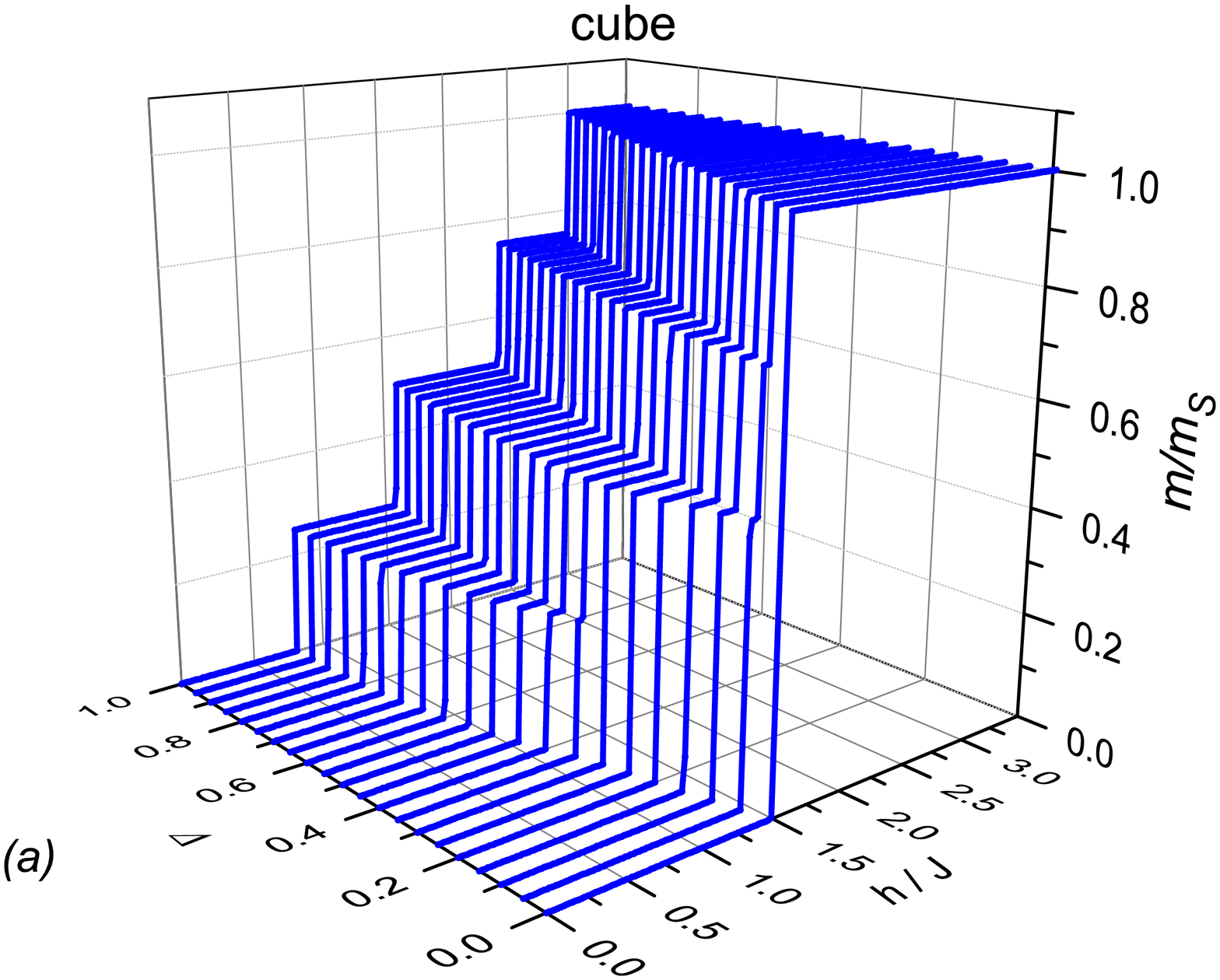}
\includegraphics[width=0.52\textwidth]{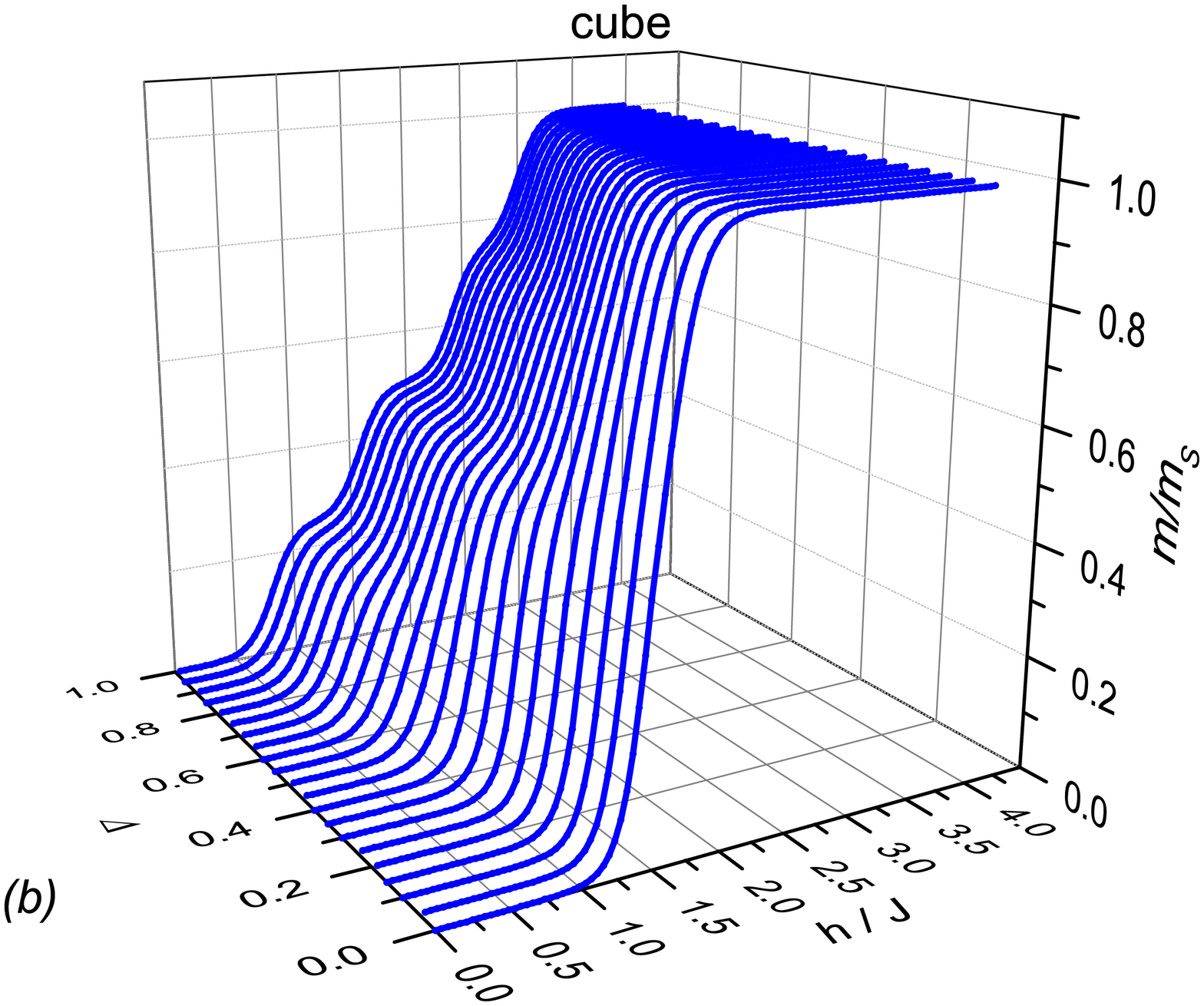}
\vspace*{-1.1cm}
\caption{The magnetization curves of the spin-$1/2$ XXZ Heisenberg cube for several values of the exchange anisotropy $\Delta$ 
and two different temperatures: (a) $k_{\rm{B}}T/J = 0.001$; (b) $k_{\rm{B}}T/J = 0.1$.}
\label{cubeM}
\end{figure}

The isothermal magnetization curves of the spin-1/2 XXZ Heisenberg cube, which are depicted in Fig.~\ref{cubeM}, provide an independent check of the aforementioned findings. The spin-1/2 Ising cube ($\Delta=0$) shows in the low-temperature magnetization curve 
just a trivial plateau at zero magnetization, whereas the spin-1/2 XXZ Heisenberg cube exhibits another two plateaux at one-half and three-quarters of the saturation magnetization for strong enough easy-axis anisotropies $0 < \Delta \lesssim 0.32$ [Fig.~\ref{cubeM}(a)]. In addition, the one-fourth plateau starts to develop in the low-temperature magnetization curve of the spin-1/2 XXZ Heisenberg cube if the anisotropy parameter $\Delta\gtrsim0.32$. Of course, the intermediate plateaux gradually diminish from the magnetization curves of the spin-1/2 XXZ Heisenberg cube at higher temperatures, whereas the moderate temperature $k_{\rm{B}}T/J = 0.1$ is sufficiently high to make all intermediate plateaux (except trivial zero plateau) indiscernible for $\Delta \lesssim 0.3$ [see Fig.~\ref{cubeM}(b)].

\begin{figure}[t]
\hspace{-0.5cm}
\includegraphics[width=0.52\textwidth]{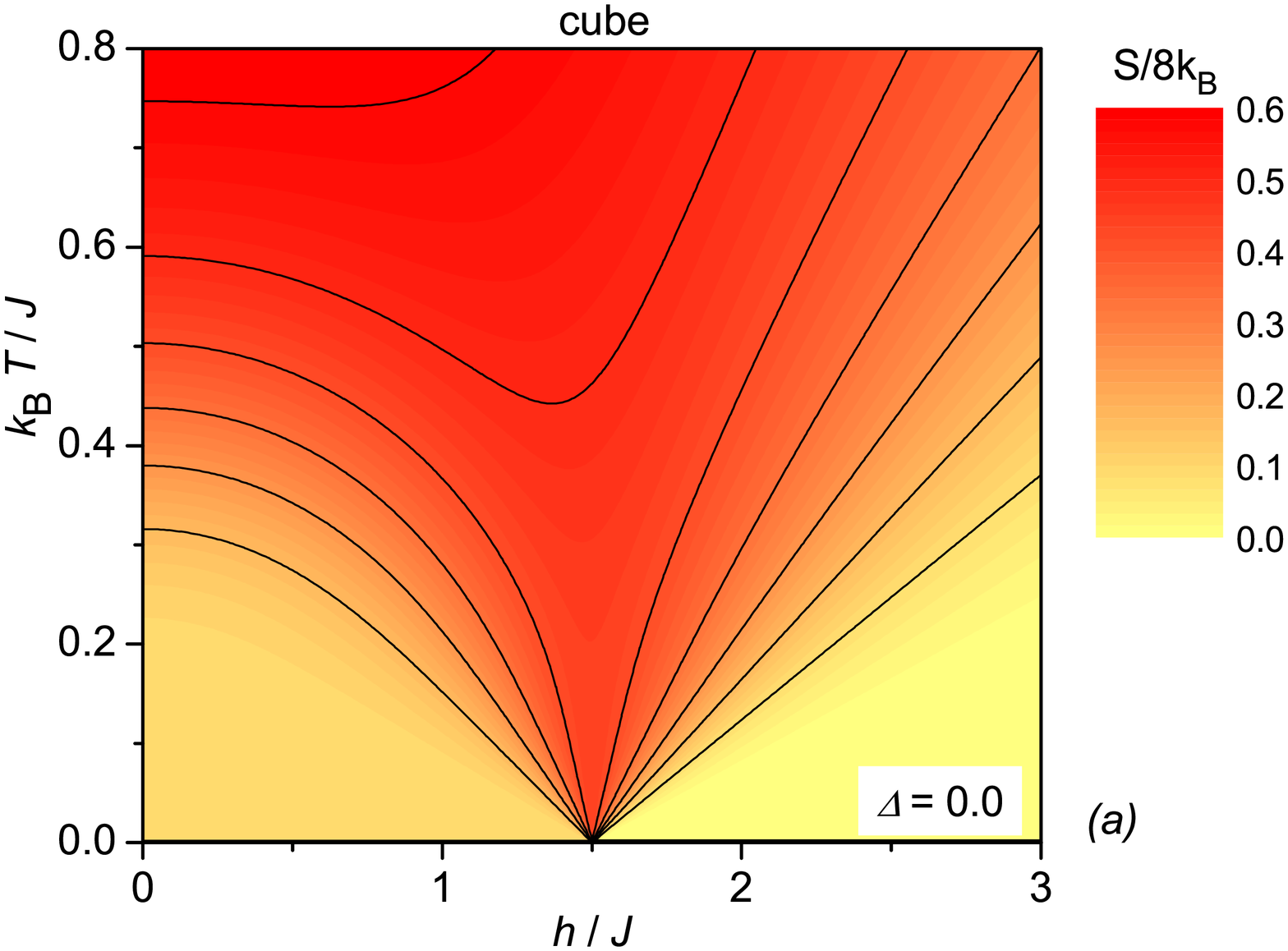}
\includegraphics[width=0.52\textwidth]{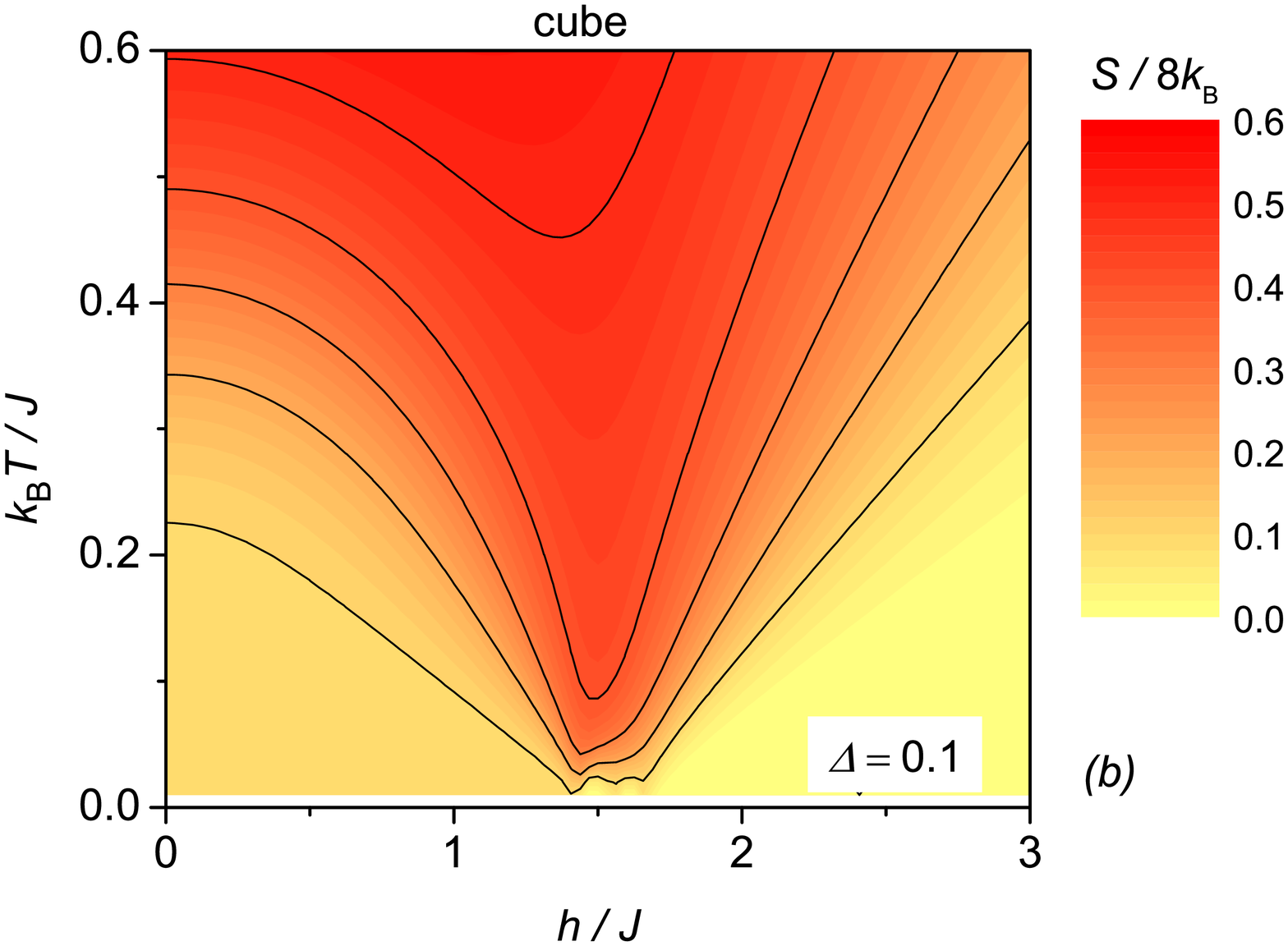}
\includegraphics[width=0.52\textwidth]{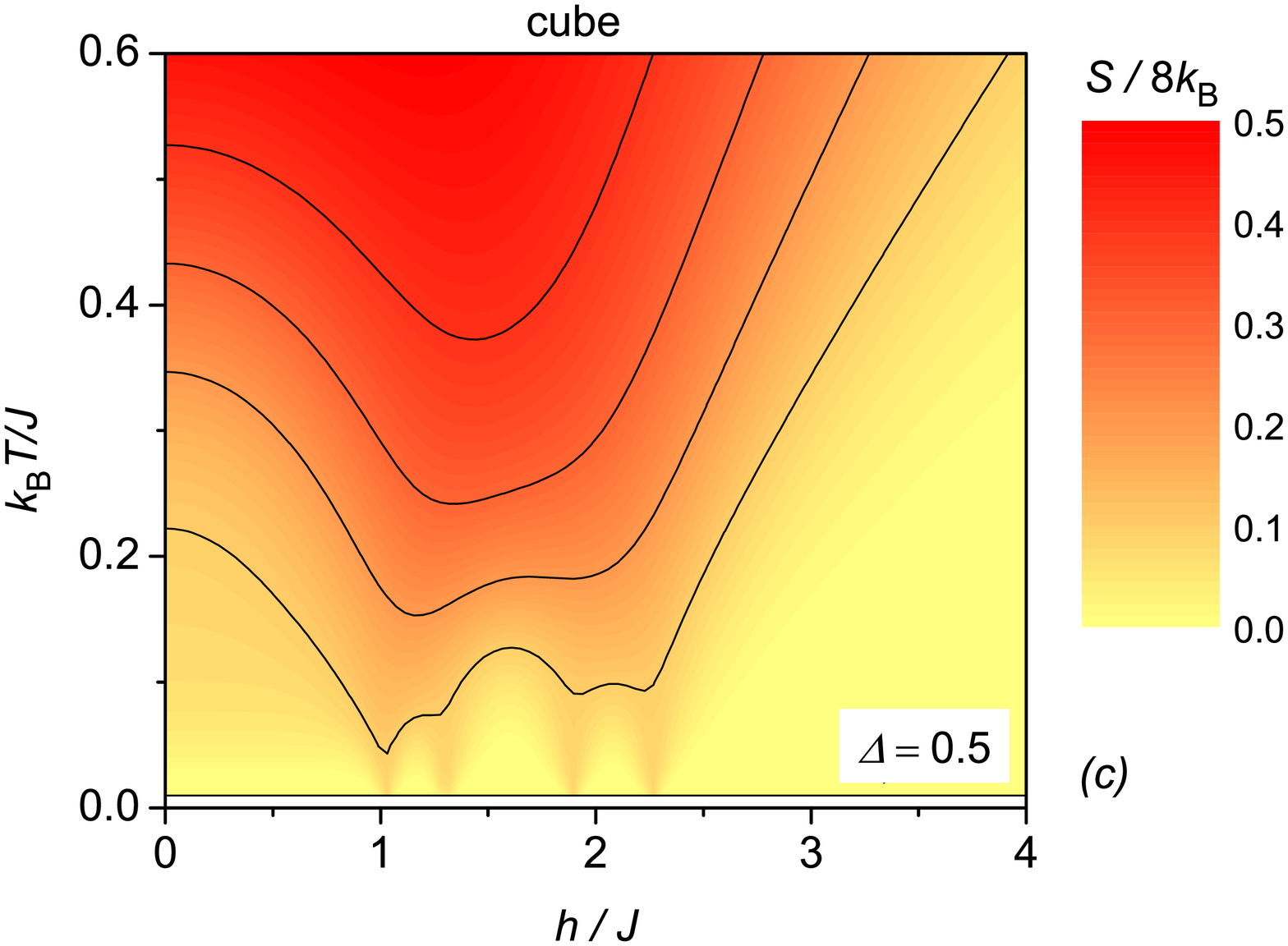}
\includegraphics[width=0.52\textwidth]{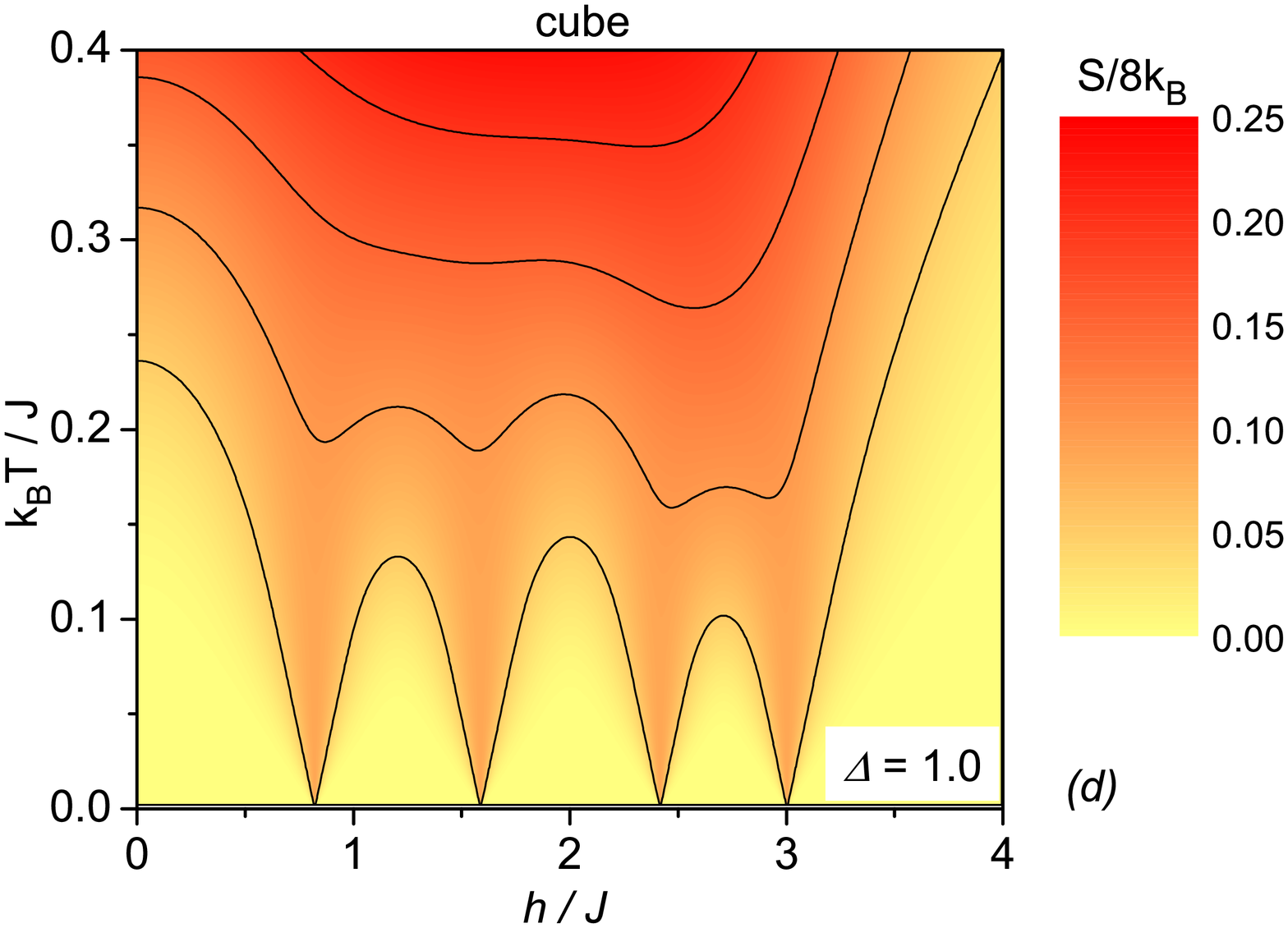}
\vspace*{-1.1cm}
\caption{A density plot of the entropy per spin of the spin-$1/2$ XXZ Heisenberg cube as a function of the magnetic field and temperature for four different values of the anisotropy parameter: (a) $\Delta=0.0$; (b) $\Delta=0.1$; (c) $\Delta=0.5$; (d) $\Delta=1.0$.}
\label{cubeE}
\end{figure}

The isentropy lines of the spin-1/2 XXZ Heisenberg cube are shown in Fig.~\ref{cubeE} in the field-temperature plane for four different values of the exchange anisotropy $\Delta$. It can be seen from Fig.~\ref{cubeE}(a) that the spin-1/2 Ising cube 
exhibits an enhanced magnetocaloric effect close to the critical field $h_c/J=3/2$, at which the abrupt magnetization jump from zero to the saturated value occurs. The qualitatively similar isentropy lines can be found for the spin-1/2 XXZ Heisenberg cube with strong enough easy-axis anisotropy $\Delta=0.1$ [see Fig.~\ref{cubeE}(b)], where the anomalous magnetocaloric response is smudged over a narrow field range around $h/J \approx 3/2$ due to three different crossing of the energy levels. Contrary to this, the spin-1/2 XXZ Heisenberg cube with a less pronounced exchange anisotropy $\Delta=0.5$ [Fig.~\ref{cubeE}(c)] or completely isotropic coupling $\Delta=1.0$ [Fig.~\ref{cubeE}(d)] 
exhibits a marked non-monotonous dependence of the isentropy lines, which follows from four distant crossings of the lowest-energy levels closely connected with the respective magnetization steps. 

\subsection{Spin-1/2 XXZ Heisenberg icosahedron}
\begin{figure}[t]
\includegraphics[width=0.52\textwidth]{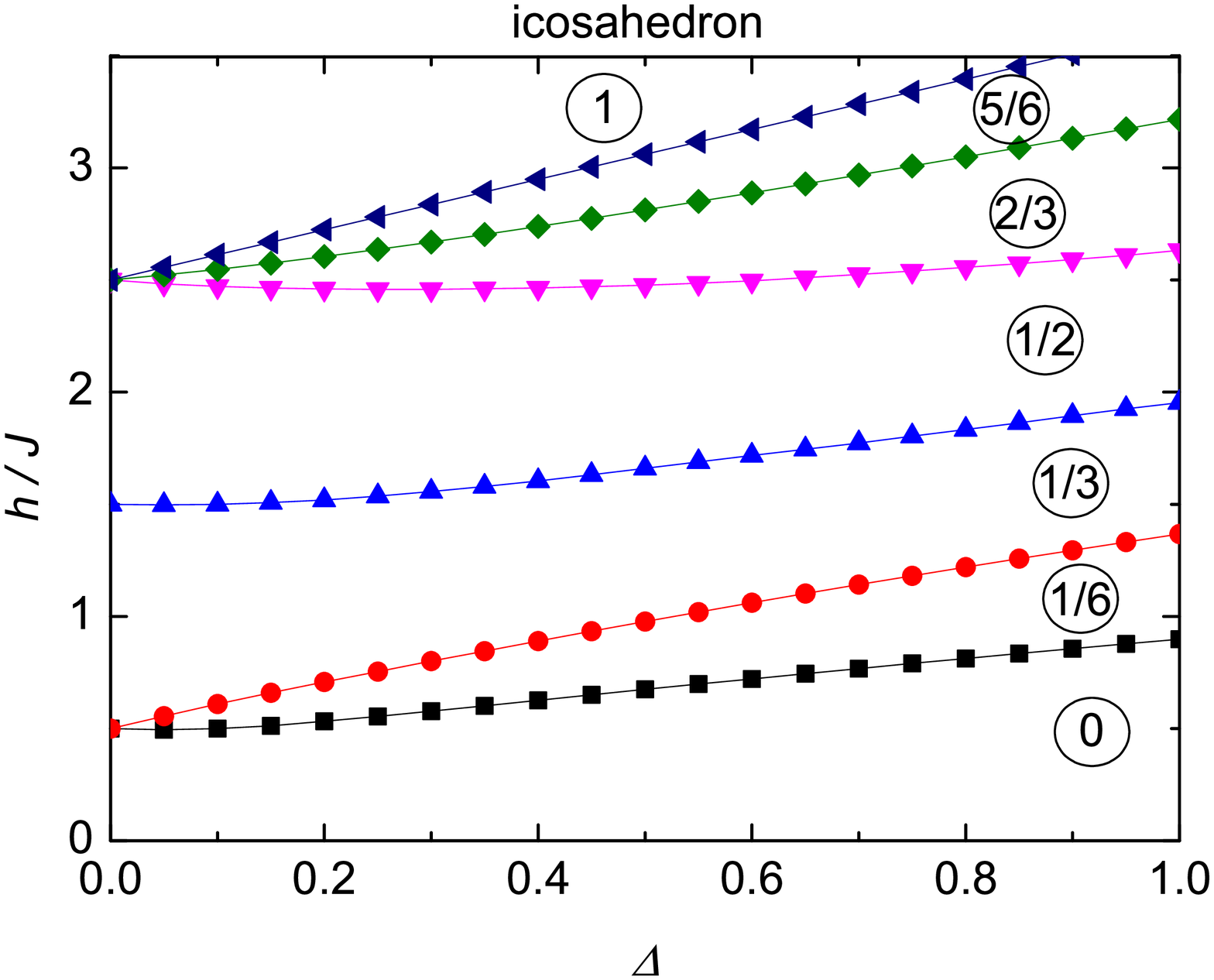}
\vspace*{-0.9cm}
\caption{The ground-state phase diagram of the spin-$1/2$ XXZ Heisenberg icosahedron in the $\Delta-h/J$ plane. 
 The acronyms
 determine the magnetization of a given lowest-energy 
eigenstate normalized with respect to its saturation value.}
\label{icoF}
\end{figure}

The ground-state phase diagram of the spin-1/2 XXZ Heisenberg icosahedron shown in Fig.~\ref{icoF} indicates a remarkable diversity of the lowest-energy eigenstates, each of which comes from the one and just one sector with available value of the total spin $S_T^z$. The most widespread lowest-energy eigenstates of the spin-1/2 XXZ Heisenberg icosahedron are the ones with the $z$-component of the total spin $S_T^z=0$, $2$ and $3$, which already appear in a zero-temperature magnetization curve of the spin-1/2 Ising icosahedron as the zero, one-third and one-half plateaux. However, the spin-1/2 XXZ Heisenberg icosahedron also exhibits for any $\Delta \neq 0$ another three lowest-energy states, which should be manifested in a zero-temperature magnetization curve as one-sixth, two-thirds and five-sixths plateaux. It could be thus concluded that two abrupt magnetization jumps and one smaller magnetization step retrieved in the Ising limit $\Delta=0$ are replaced through five smaller magnetization steps on assumption that $\Delta \neq 0$.

\begin{figure}[t]
\includegraphics[width=0.52\textwidth]{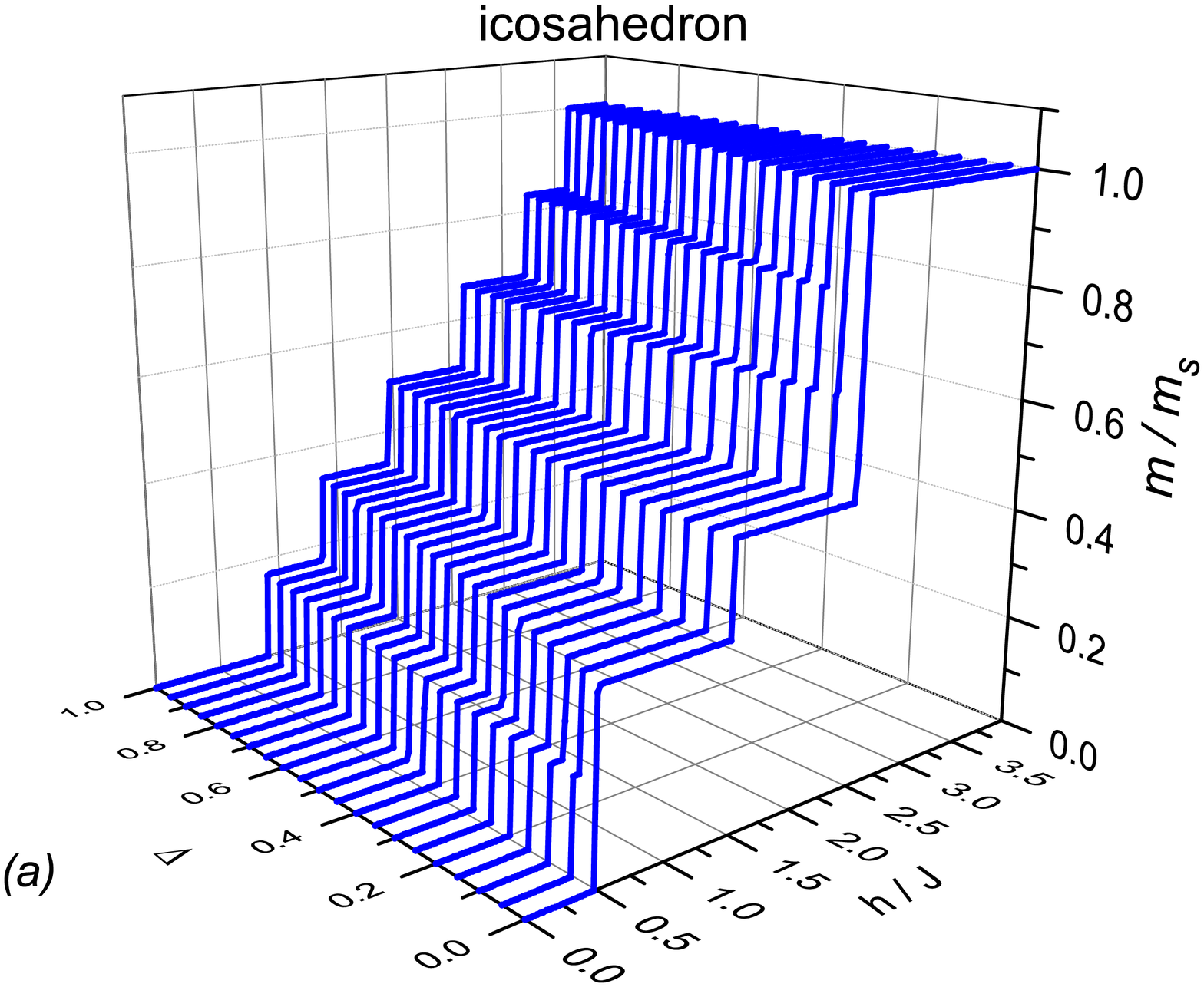}
\includegraphics[width=0.52\textwidth]{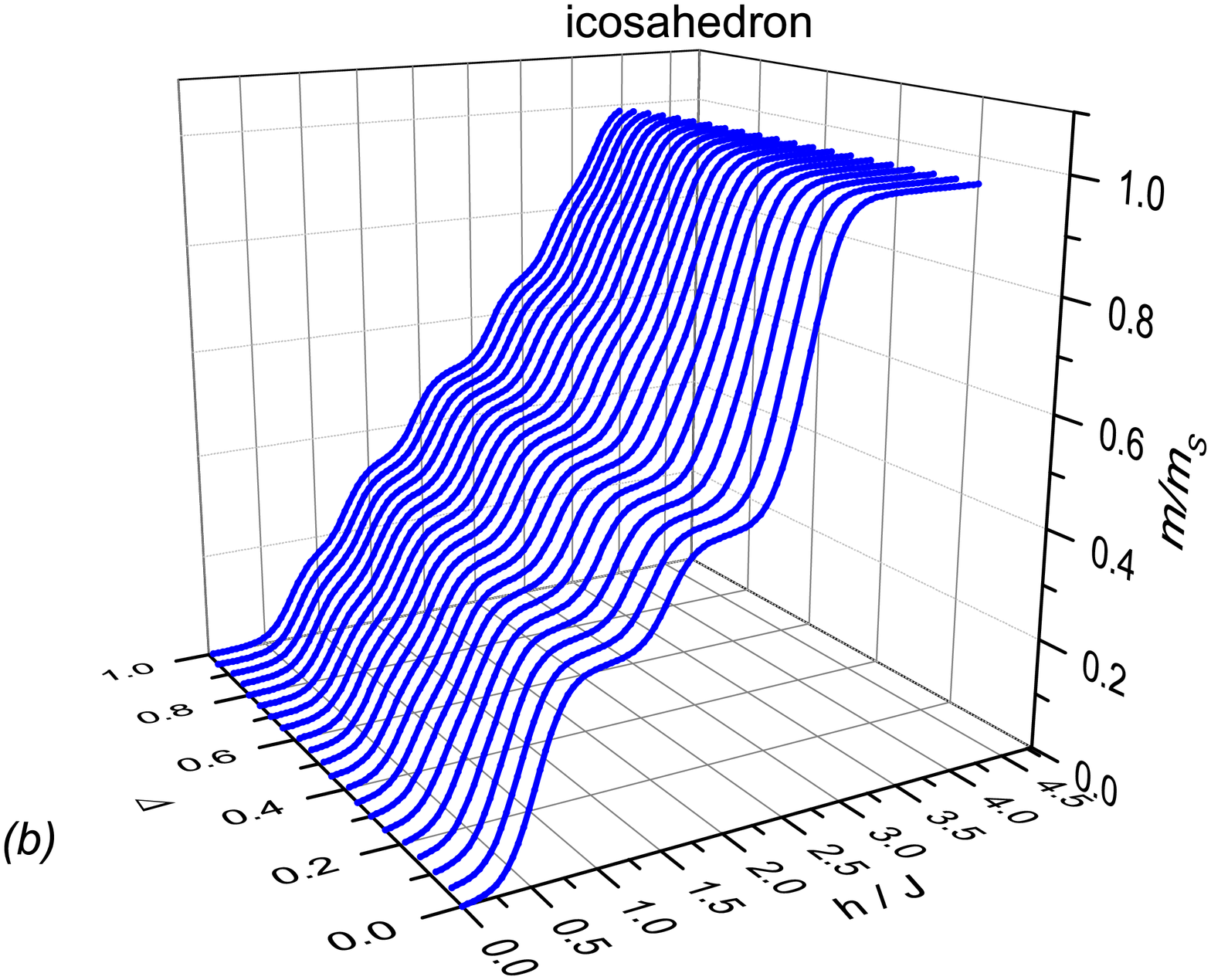}
\vspace*{-1.1cm}
\caption{The magnetization curves of the spin-$1/2$ XXZ Heisenberg icosahedron for several values of the exchange anisotropy $\Delta$ 
and two different temperatures: (a) $k_{\rm{B}}T/J = 0.001$; (b) $k_{\rm{B}}T/J = 0.1$.}
\label{icoM}
\end{figure}

To verify this issue, we have plotted in Fig.~\ref{icoM}(a) the isothermal magnetization curves of the spin-1/2 XXZ Heisenberg icosahedron at sufficiently low temperature $k_{\rm{B}}T/J = 0.001$. In agreement with our expectation, the spin-1/2 Ising icosahedron 
exhibits three magnetization plateaux at zero, one-third and one-half of the saturation magnetization, whereas the spin-1/2 XXZ Heisenberg icosahedron with $\Delta \neq 0$ additionally displays three novel magnetization plateaux at one-sixth, two-thirds and five-sixths of the saturation magnetization. It is quite interesting to notice that the intermediate one-third and one-half magnetization plateaux become narrower and zero magnetization plateau contrarily wider upon strengthening of $\Delta$. The effect of increasing temperature on the magnetization curves of the spin-1/2 XXZ Heisenberg icosahedron is illustrated in Fig.~\ref{icoM}(b). It can be seen from this figure that only three widest magnetization plateaux at zero, one-third and one-half of the saturation magnetization are still clearly visible at the moderate temperature $k_{\rm{B}}T/J = 0.1$ for any $\Delta$, while other narrower plateaux are merely reflected by less pronounced inflection points.

\begin{figure}[t]
\includegraphics[width=0.52\textwidth]{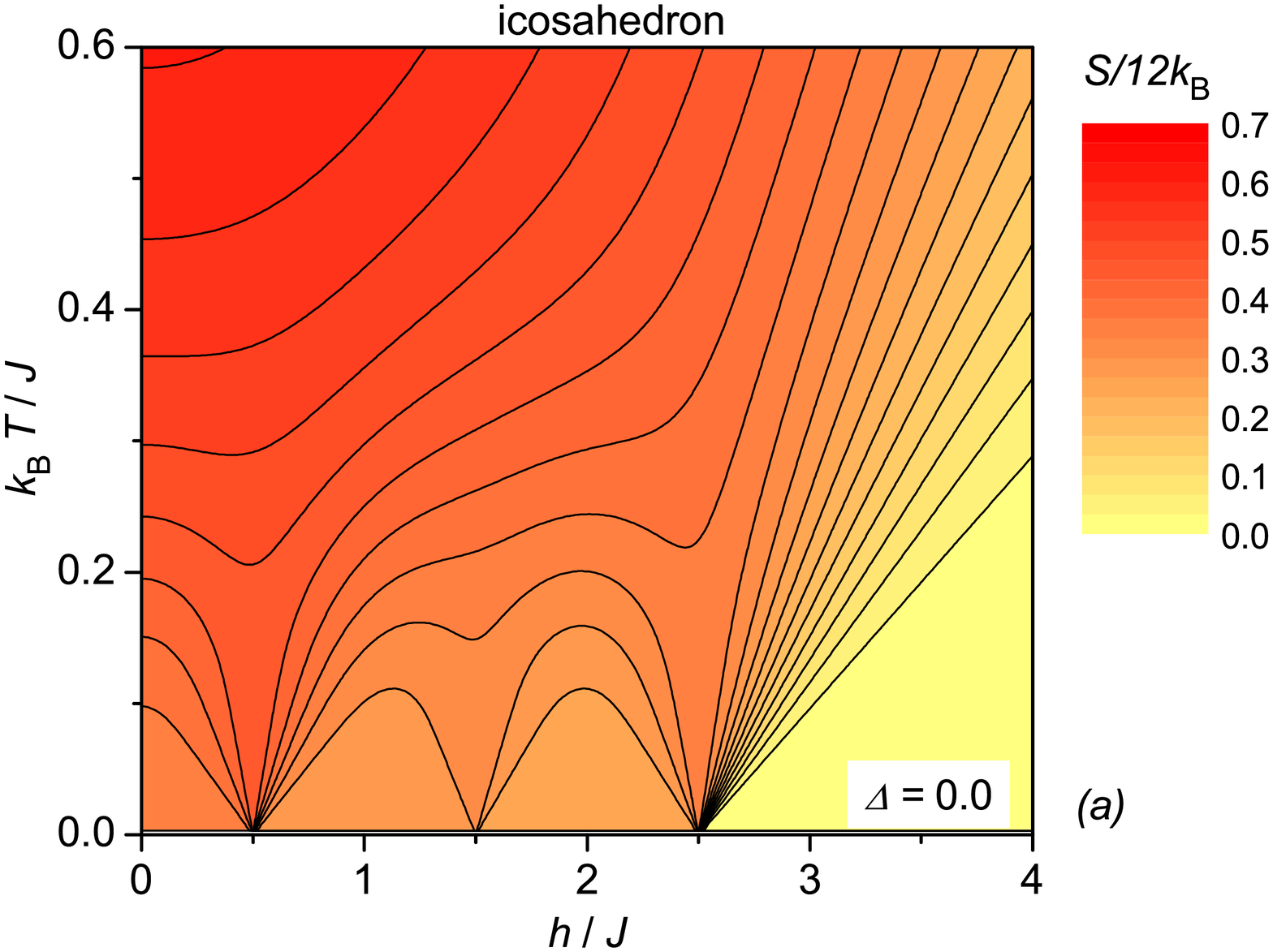}
\includegraphics[width=0.52\textwidth]{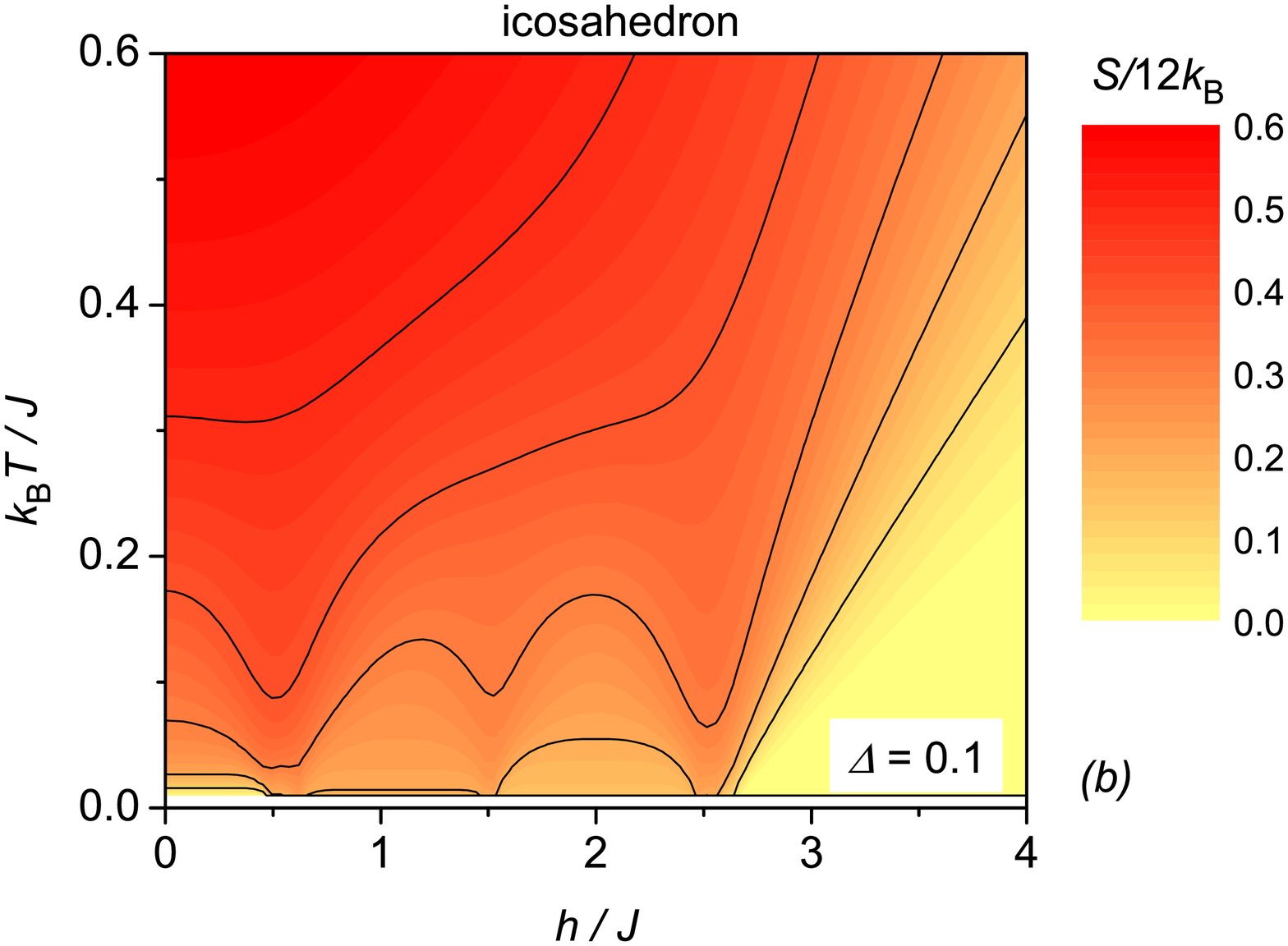}
\includegraphics[width=0.52\textwidth]{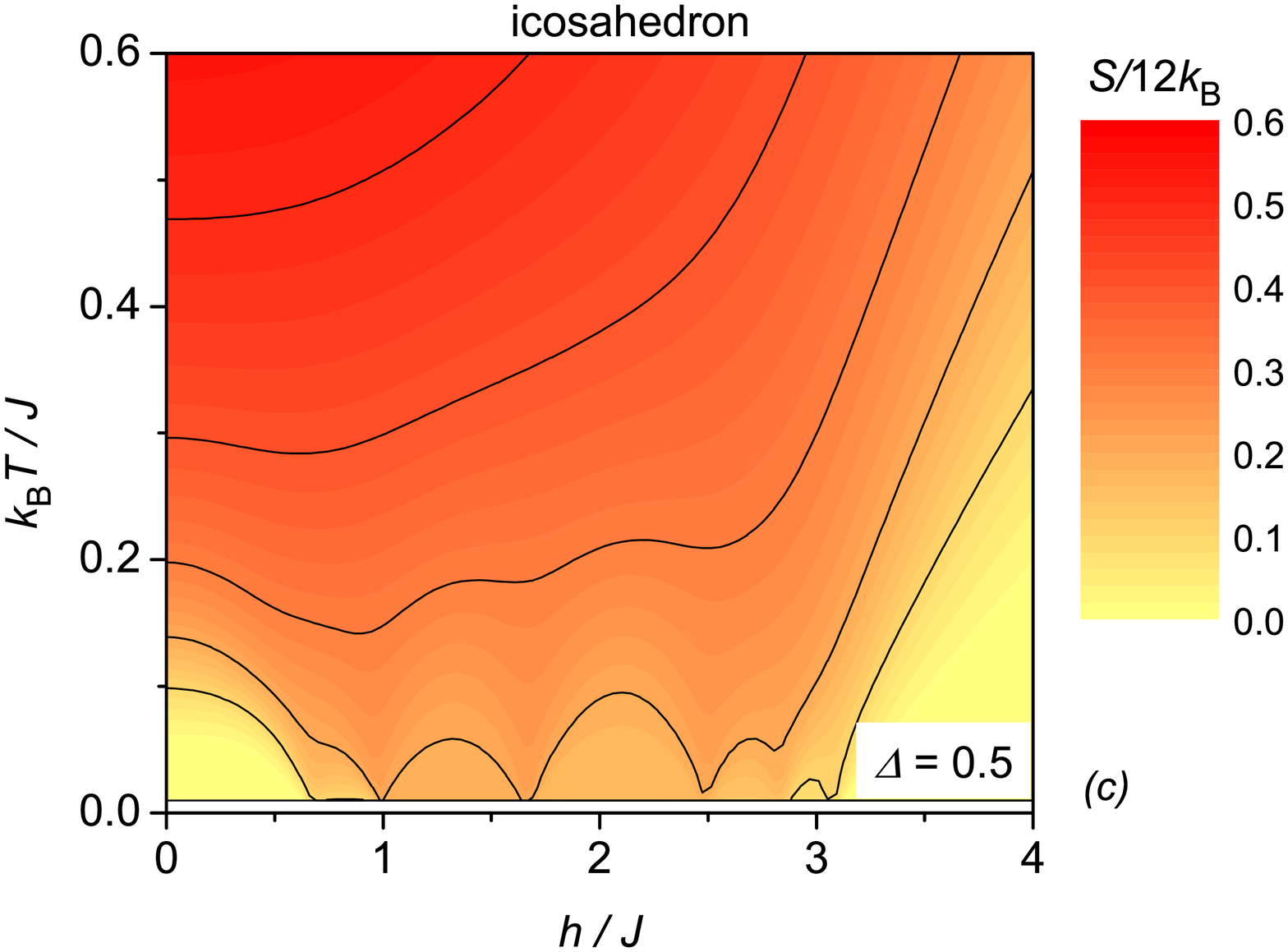}
\includegraphics[width=0.52\textwidth]{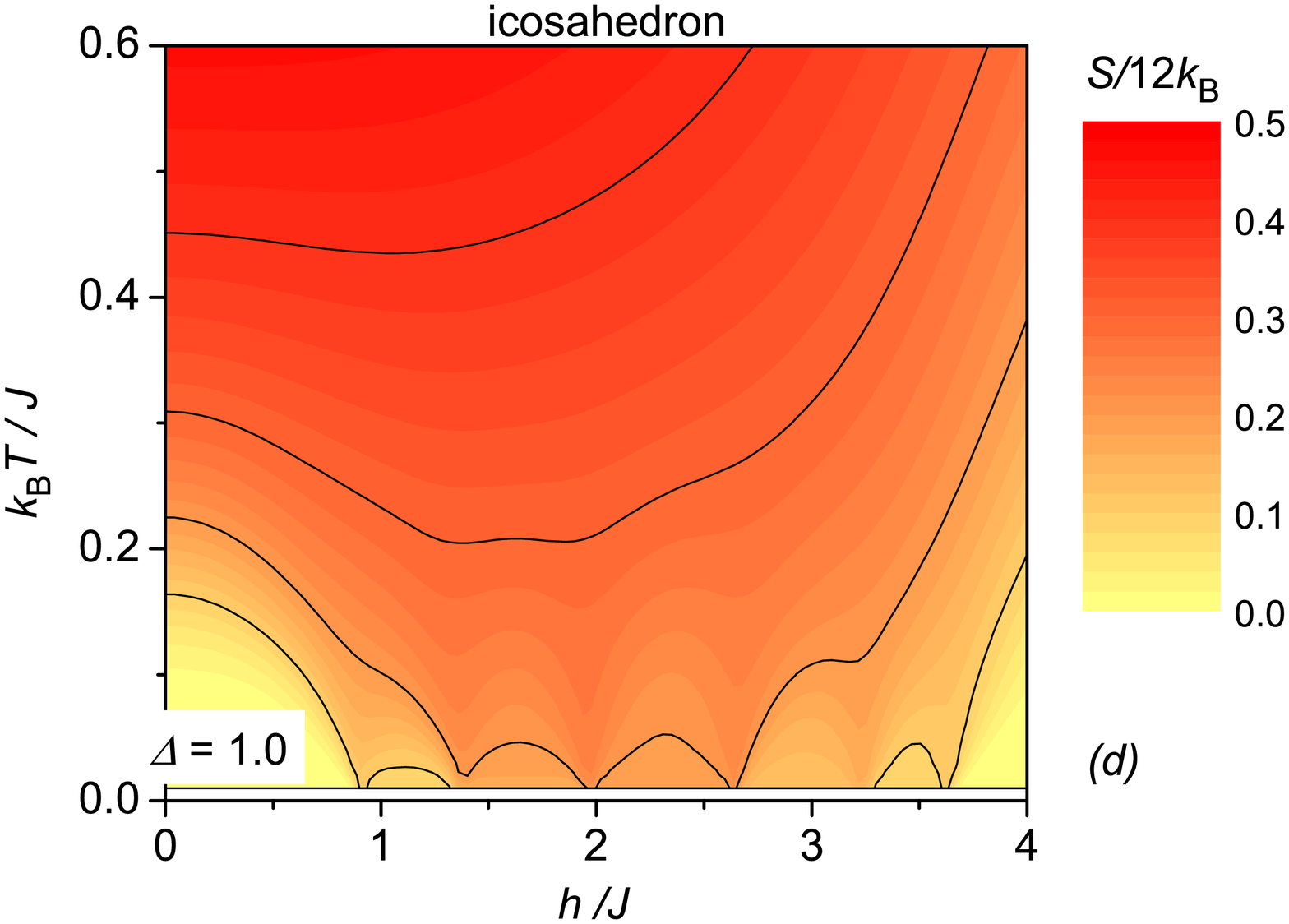}
\vspace*{-1.1cm}
\caption{A density plot of the entropy per spin of the spin-$1/2$ XXZ Heisenberg icosahedron as a function of the magnetic field and temperature for four different values of the anisotropy parameter: (a) $\Delta=0.0$; (b) $\Delta=0.1$; (c) $\Delta=0.5$; (d) $\Delta=1.0$.}
\label{icoeE}
\end{figure}

Furthermore, it is obvious from Fig.~\ref{icoeE} that the magnetocaloric response of the spin-1/2 XXZ Heisenberg icosahedron basically depends on a choice of the exchange anisotropy $\Delta$. The spin-1/2 Ising icosahedron ($\Delta=0.0$) and the spin-1/2 XXZ Heisenberg icosahedron with an extremely high easy-axis anisotropy ($\Delta=0.1$) generally show an enhanced magnetocaloric effect around three different magnetic fields [see Fig.~\ref{icoeE}(a)-(b)]. For comparison, the spin-1/2 XXZ Heisenberg icosahedron either with the moderate easy-axis anisotropy ($\Delta=0.5$) or the completely isotropic coupling ($\Delta=1.0$) 
exhibits during the adiabatic demagnetization an abrupt change of the temperature in the vicinity of all six level-crossing fields.

\subsection{Spin-1/2 XXZ Heisenberg dodecahedron}
\begin{figure}[t]
\includegraphics[width=0.52\textwidth]{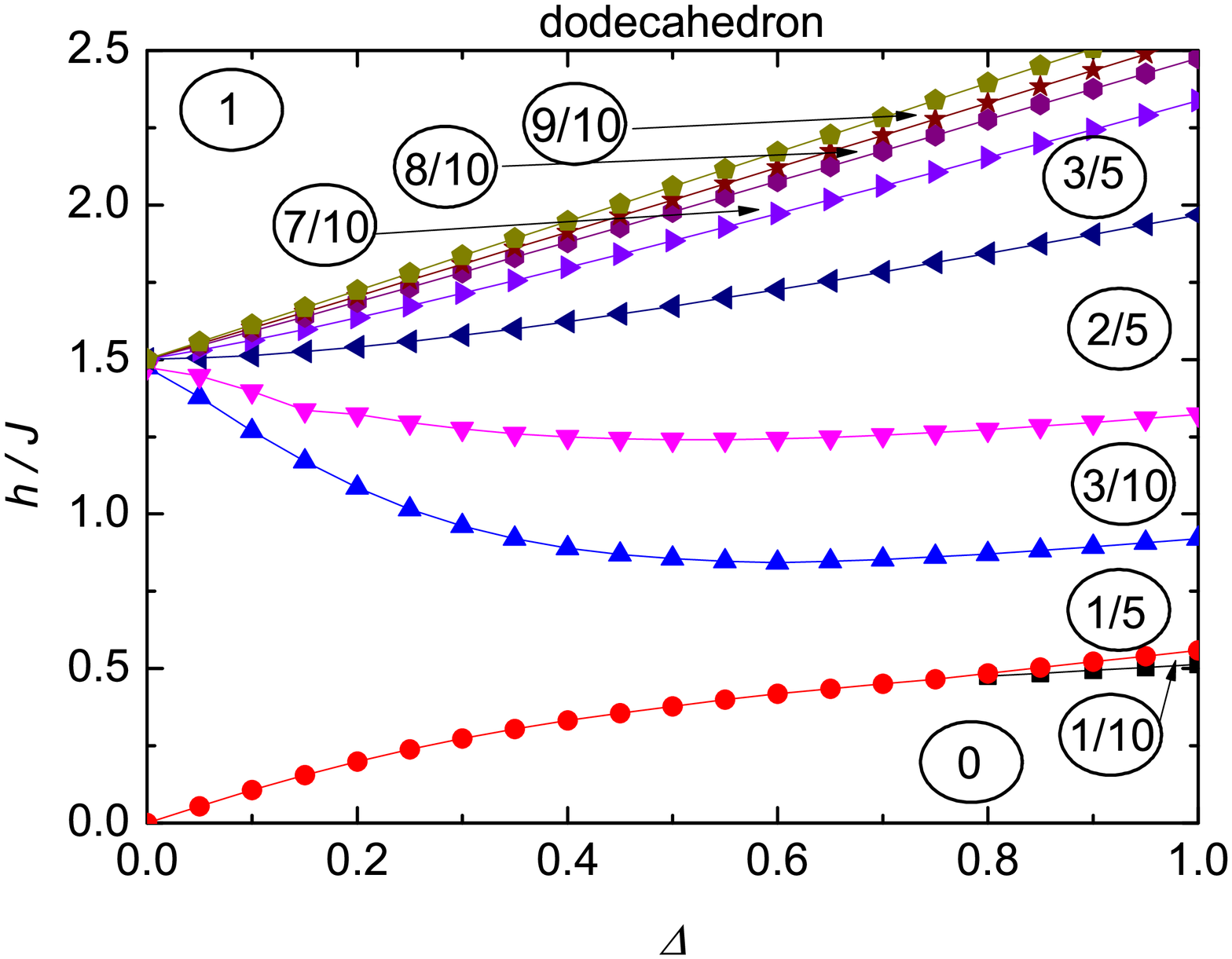}
\vspace*{-0.9cm}
\caption{(a) The ground-state phase diagram of the spin-$1/2$ XXZ Heisenberg dodecahedron in the $\Delta-h/J$ plane.
The acronyms 
 determine the magnetization of a given lowest-energy 
eigenstate normalized with respect to its saturation value.}
\label{dodeF}
\end{figure}

Last but not least, the ground-state phase diagram of the spin-1/2 XXZ Heisenberg dodecahedron is depicted in Fig.~\ref{dodeF} in the $\Delta-h/J$ plane. It can be understood from Fig.~\ref{dodeF} that the spin-1/2 Ising dodecahedron should 
exhibit in the zero-temperature magnetization curve two noticeable magnetization jumps at zero field and the saturation field, which determine rise and fall of the one-fifth magnetization plateau, respectively. Contrary to this, the spin-1/2 XXZ Heisenberg dodecahedron with $\Delta \neq 0$ may additionally display up to eight other lowest-energy eigenstates, which become evident in 
the zero-temperature magnetization curve as fractional plateaux at all integer fractions of the number ten with exception of one-half. Another striking feature is that the one-tenth plateau is realized as the lowest-energy state of the spin-1/2 XXZ Heisenberg dodecahedron just at relatively weak easy-axis anisotropies $\Delta \gtrsim 0.8$, while this plateau is completely absent at stronger easy-axis anisotropies $\Delta \lesssim 0.8$ quite similarly as the ever absent one-half plateau. It could be thus concluded that the spin-1/2 XXZ Heisenberg dodecahedron shows besides a few smaller magnetization steps either one or two magnetization jumps depending on whether $\Delta \lesssim 0.8$ or $\Delta \gtrsim 0.8$, respectively. 

\begin{figure}[t]
\includegraphics[width=0.52\textwidth]{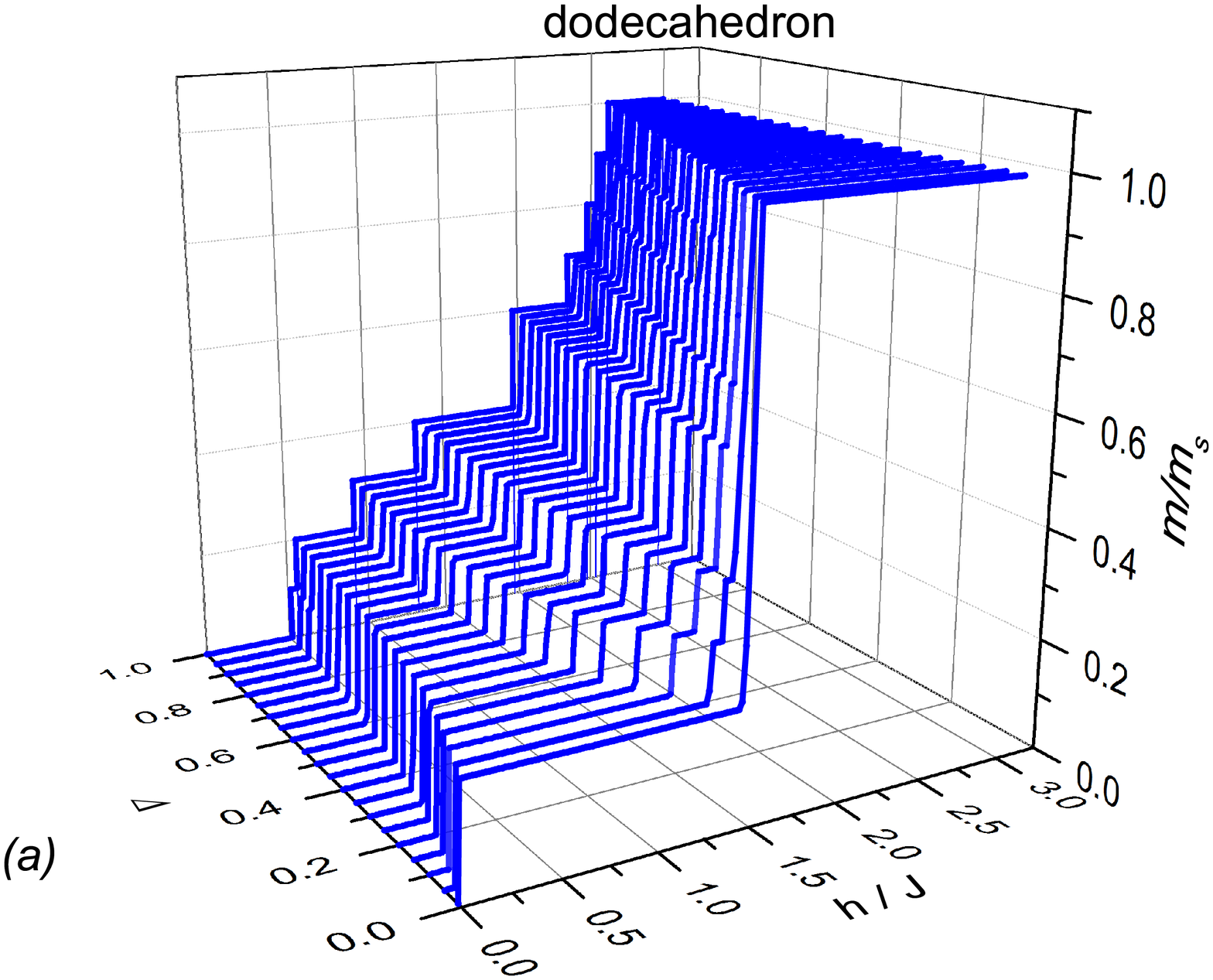}
\includegraphics[width=0.52\textwidth]{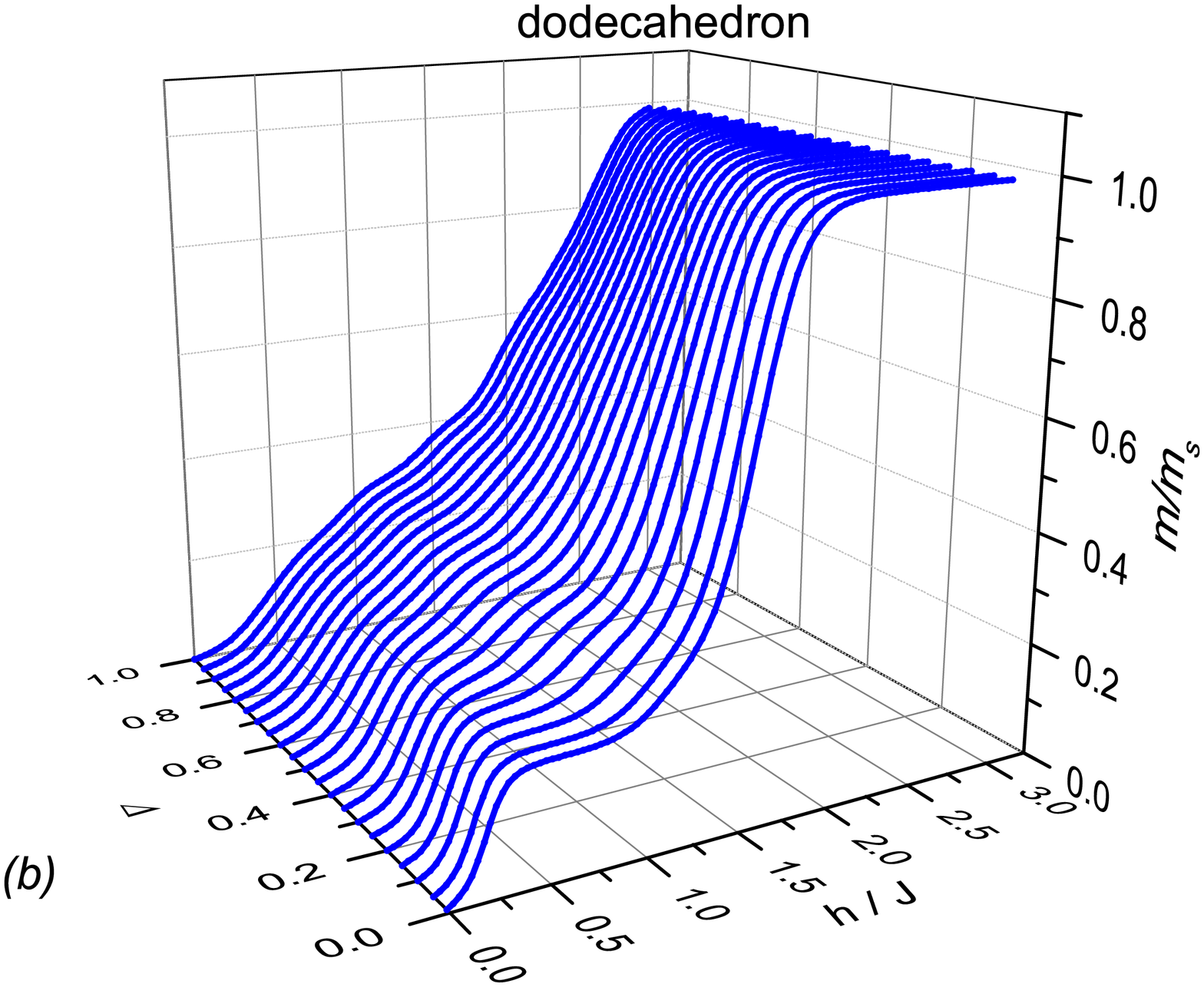}
\vspace*{-1.1cm}
\caption{The magnetization curves of the spin-$1/2$ XXZ Heisenberg dodecahedron for several values of the exchange anisotropy $\Delta$ 
and two different temperatures: (a) $k_{\rm{B}}T/J = 0.001$; (b) $k_{\rm{B}}T/J = 0.1$.}
\label{dodeM}
\end{figure}

The aforementioned findings can be corroborated by the isothermal magnetization curves, which are plotted in Fig.~\ref{dodeM}(a) for the spin-1/2 XXZ Heisenberg dodecahedron at sufficiently low temperature $k_{\rm{B}}T/J = 0.001$. It actually turns out that the wide one-fifth magnetization plateau observable in the Ising limit ($\Delta=0$) is substantially reduced upon switching on the quantum $xy$-part of the XXZ exchange interaction due to uprise of the novel intermediate magnetization plateaux. In a consequence of that, the one-fifth magnetization plateau of the isotropic Heisenberg dodecahedron ($\Delta=1$) is approximately by $80\%$ shorter in comparison with that of the Ising dodecahedron ($\Delta=0$) 
if the length of the relevant plateau is scaled with respect to the saturation field. The isothermal magnetization curves of the spin-1/2 XXZ Heisenberg dodecahedron shown in Fig.~\ref{dodeM}(b) shed light on the effect of rising temperature. Apparently, the temperature-driven smoothing of the stepwise magnetization curve is already sufficient at the moderate temperature $k_{\rm{B}}T/J = 0.1$ in order to suppress all intermediate plateaux except the widest one-fifth plateau, which persists at extremely strong exchange anisotropies $\Delta \lesssim 0.3$ of the easy-axis type. 

\begin{figure}[t]
\includegraphics[width=0.52\textwidth]{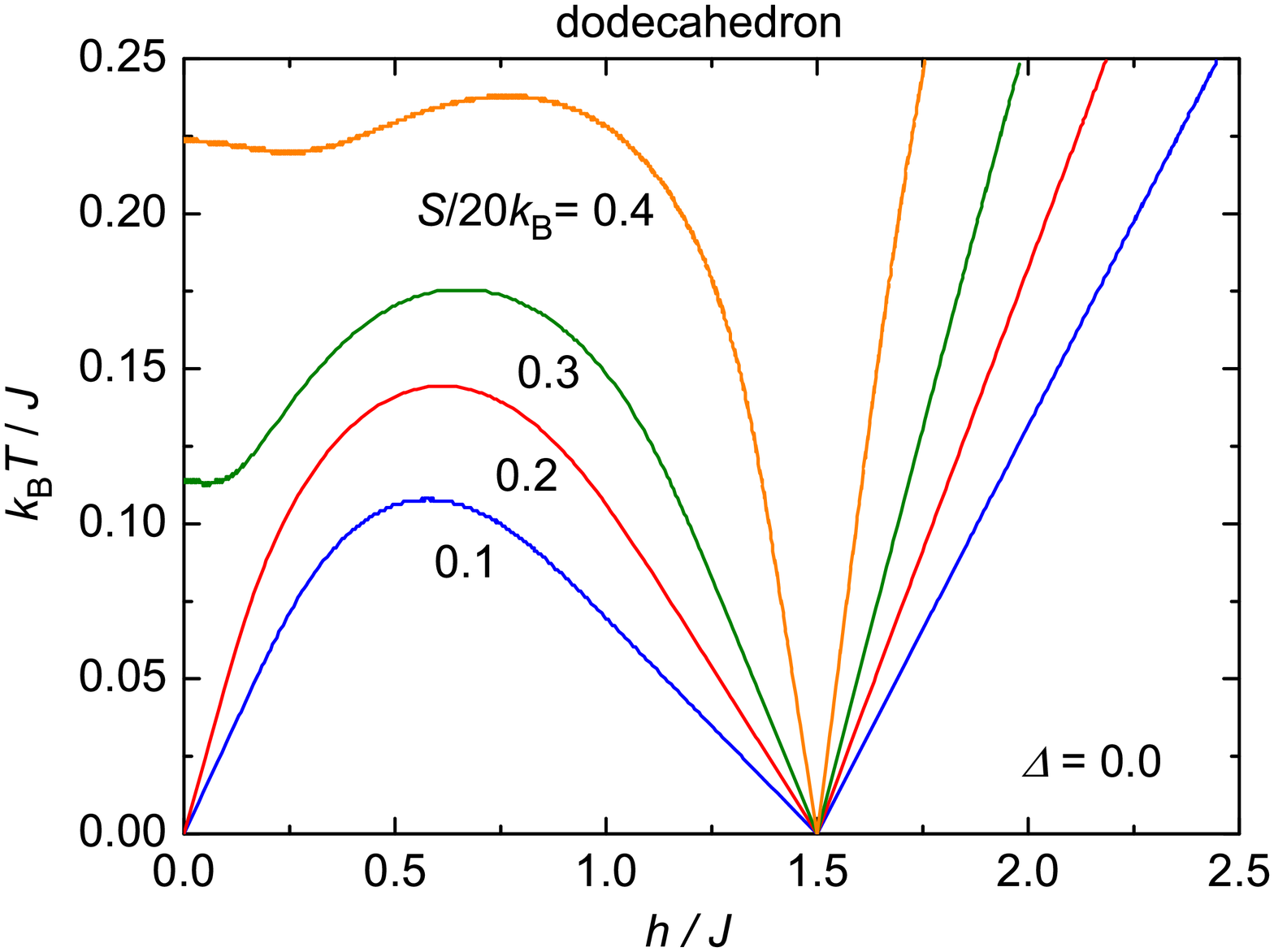}
\includegraphics[width=0.52\textwidth]{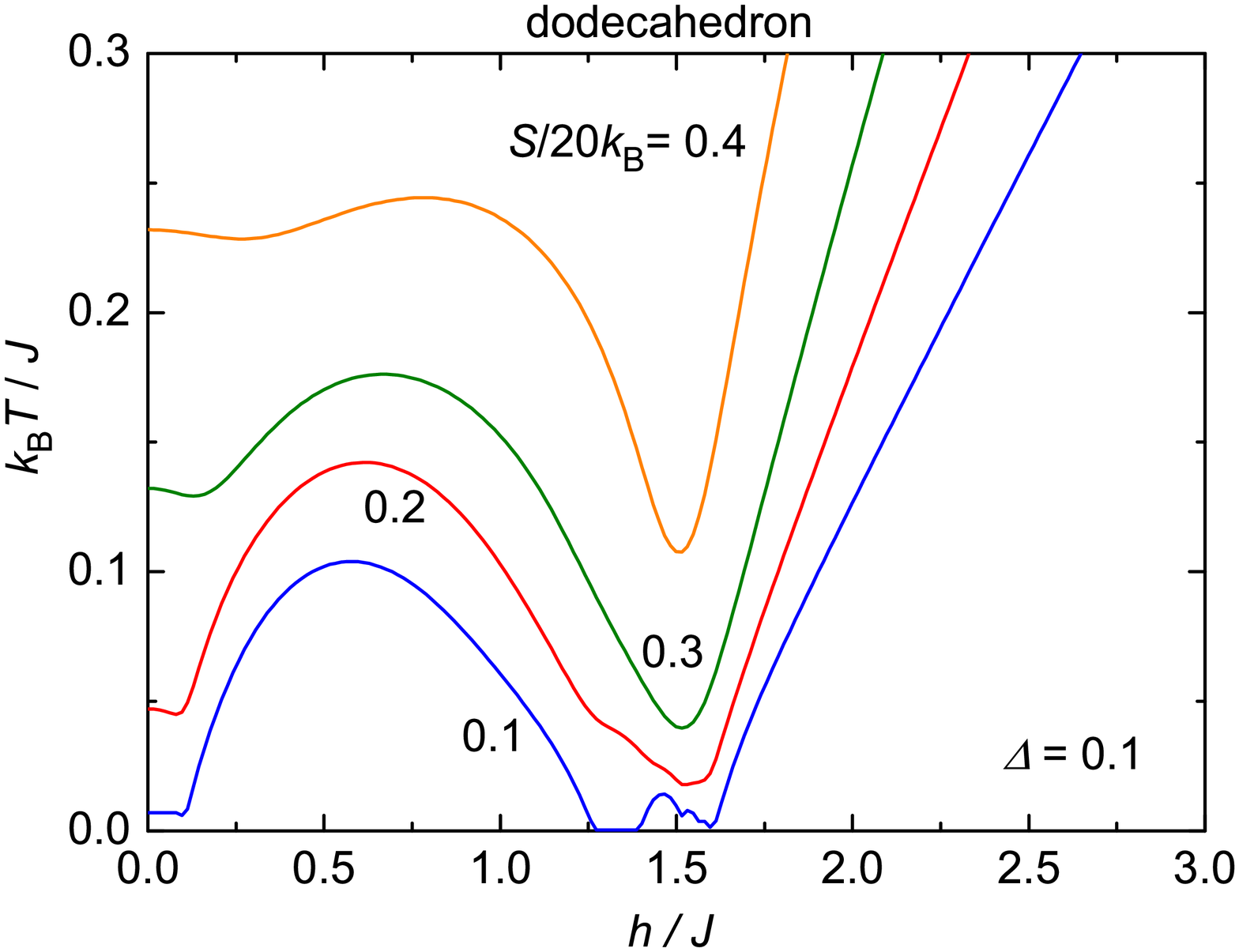}
\includegraphics[width=0.52\textwidth]{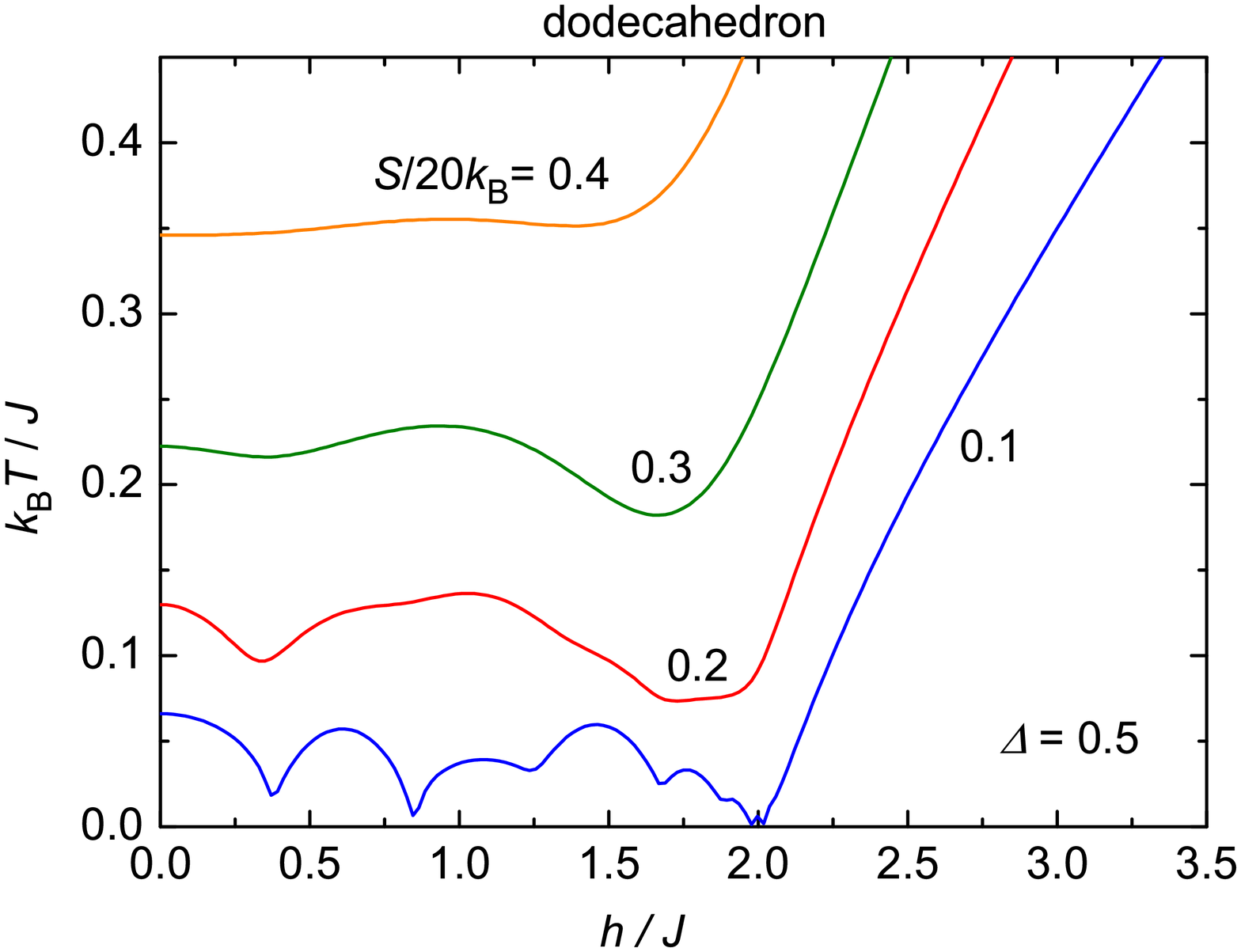}
\includegraphics[width=0.52\textwidth]{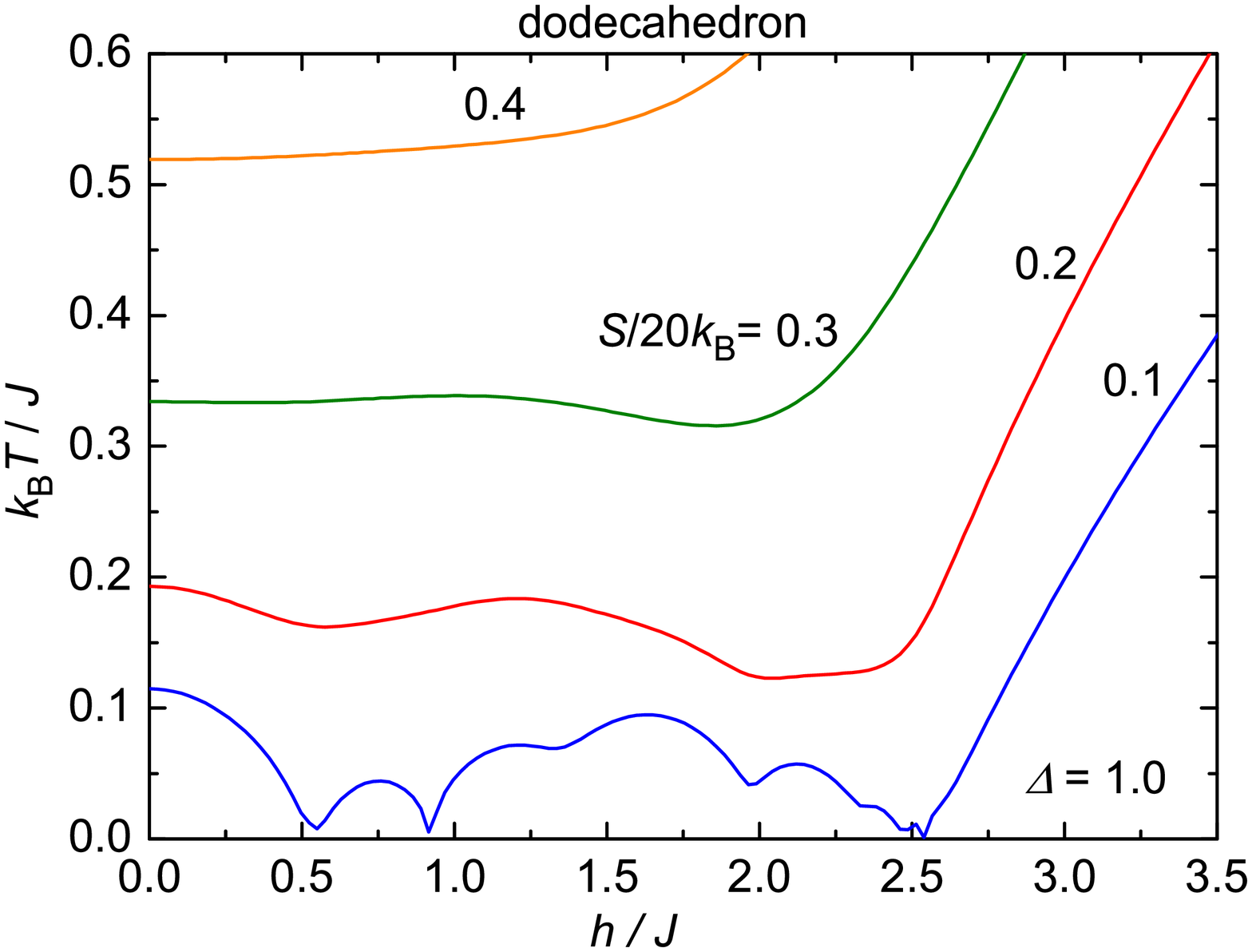}
\vspace*{-1.1cm}
\caption{A few isentropy lines of the spin-$1/2$ XXZ Heisenberg dodecahedron in the field-temperature plane for four different values of the anisotropy parameter: (a) $\Delta=0.0$; (b) $\Delta=0.1$; (c) $\Delta=0.5$; (d )$\Delta=1.0$.}
\label{dodeE}
\end{figure}

For completeness, Fig.~\ref{dodeE} illustrates a few typical isentropy lines of the spin-1/2 XXZ Heisenberg dodecahedron for four different values of the exchange anisotropy $\Delta$, which bring insight into the magnetocaloric response of temperature achieved upon adiabatic variation of the magnetic field. The absence of zero magnetization plateau in a magnetization process of the spin-1/2 Ising dodecahedron [Fig.~\ref{dodeE}(a)] repeatedly causes a giant magnetocaloric effect in a proximity of zero magnetic field, which makes from an experimental realization of the spin-1/2 Ising dodecahedron promising refrigerant quite similarly as an experimental realization of the spin-1/2 Ising octahedron is \cite{stre15}. However, it turns out that arbitrary but non-zero quantum $xy$-part of the XXZ exchange interaction regrettably suppresses this spectacular magnetocaloric feature. For instance, the spin-1/2 XXZ Heisenberg dodecahedron with the extremely strong easy-axis anisotropy shows an enhanced magnetocaloric effect upon attenuation of the magnetic field, which is successively followed by unfavourable uprise of temperature (albeit in a restricted field range) as the magnetic field is removed adiabatically (see Fig.~\ref{dodeE}(b) for $\Delta=0.1$). The adiabatic demagnetization of the spin-1/2 XXZ Heisenberg dodecahedron with a less pronounced exchange anisotropy ($\Delta=0.5$) or completely isotropic coupling ($\Delta=1.0$) are completely irrelevant for cooling purposes due to the substantial increase of temperature observable in a low-field region inherent to the zero magnetization plateau [see Fig.~\ref{dodeE}(c)-(d)]. 

\section{Conclusion}
\label{conclusion}

The present work deals with magnetic properties of the antiferromagnetic spin-1/2 XXZ Heisenberg regular polyhedra, which are examined by the use of exact analytical or numerical diagonalization method. In particular, we have clarified the question how the ground-state phase diagram, magnetization process and magnetocaloric effect of the high-symmetry quantum spin clusters evolve with the exchange anisotropy that controls a relative strength of quantum fluctuations. It has been demonstrated that the spin-1/2 XXZ Heisenberg regular polyhedra generally exhibit in the low-temperature magnetization curves greater number of the intermediate magnetization plateaux than their analogous Ising counterparts. It is quite apparent that the underlying mechanism for a formation of the novel magnetization plateaux must be of a purely quantum origin in contrast with the formation mechanism of remaining intermediate plateaux, which emerge in the semi-classical Ising limit on account of a geometric spin frustration. While the most of novel quantized magnetization plateaux appear by assuming arbitrary but non-zero $xy$-part of the XXZ exchange interaction ($\Delta \neq 0$), this need not be always the case as evidenced by two particular examples of the one-fourth plateau of the spin-1/2 XXZ Heisenberg cube or the one-tenth plateau of the spin-1/2 XXZ Heisenberg dodecahedron. 

The spin-1/2 XXZ Heisenberg regular polyhedra display a remarkable diversity of the low-temperature magnetization curves closely connected to the magnetization steps or jumps, which take place at each field-driven crossing of the lowest-energy levels corresponding to adjacent magnetization plateaux. The magnetization steps and jumps are always accompanied with an enhanced magnetocaloric effect, which causes a relatively rapid drop (rise) of temperature just above (below) of a level-crossing field when the magnetic field is removed adiabatically. It has been convincingly evidenced that arbitrary but non-zero $xy$-part of the XXZ exchange interaction regrettably suppresses a giant magnetocaloric effect, which has been recently reported for the adiabatic demagnetization of the spin-1/2 Ising octahedron, dodecahedron \cite{stre15} and cuboctahedron \cite{karl17}. The question whether or not some quantum Heisenberg spin cluster may exhibit a giant magnetocaloric effect in a proximity of zero magnetic field (due to absence of zero magnetization plateau) thus remains open.

\section*{Acknowledgments}
This work was financially supported by the grants of The Ministry of Education, Science, Research and Sport of the 
Slovak Republic under the contract Nos. VEGA 1/0331/15 and VEGA 1/0043/16, 
the Slovak Research and Development Agency provided under contract No. APVV-14-0073, 
as well as, Faculty of Science of P. J. \v{S}af\'arik University provided under contract No. VVGS-PF-2016-72606.
The full diagonalization for the dodecahedron was performed
with J. Schulenburg's \textit{spinpack}, 
see http://www-e.uni-magdeburg.de/jschulen/spin/.

\end{document}